\documentclass{article}

\usepackage{arxiv}

\usepackage[utf8]{inputenc} % allow utf-8 input
\usepackage[T1]{fontenc}    % use 8-bit T1 fonts
\usepackage{hyperref}       % hyperlinks
\usepackage{url}            % simple URL typesetting
\usepackage{booktabs}       % professional-quality tables
\usepackage{amsfonts}       % blackboard math symbols
\usepackage{nicefrac}       % compact symbols for 1/2, etc.
\usepackage{microtype}      % microtypography
\usepackage{cleveref}       % smart cross-referencing
\usepackage{graphicx}
\usepackage[numbers,square]{natbib}
\usepackage{doi}
\usepackage{multirow}
\usepackage{graphicx}
\usepackage[table,xcdraw]{xcolor}
\usepackage{float}

\title{LLM-Agent-UMF: LLM-based Agent Unified Modeling Framework for Seamless Design of Multi Active/Passive Core-Agent Architectures}

\date{}

\usepackage{authblk}

\newbox{\orcid}\sbox{\orcid}{\includegraphics[scale=0.06]{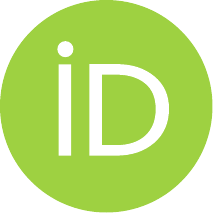}}

\author[1,2,3]{%
	\href{https://orcid.org/0009-0005-8915-7905}{\usebox{\orcid}\hspace{1mm}Amine Ben~Hassouna\thanks{\texttt{amine.benhassouna@medtech.tn, amine.benhassouna@dracodes.com (Corresponding author)}}}%
}

\author[1,2]{%
	\href{https://orcid.org/0009-0002-2629-5102}{\usebox{\orcid}\hspace{1mm}Hana Chaari\thanks{\texttt{hana.chaari@medtech.tn, hana.chaari@dracodes.com}}}\textsuperscript{\textsection}%
}

\author[1,2]{%
	\href{https://orcid.org/0009-0008-3435-740X}{\usebox{\orcid}\hspace{1mm}Ines Belhaj\thanks{\texttt{ines.bel-hadj@medtech.tn, ines.bel-hadj@dracodes.com}}}\textsuperscript{\textsection}%
}

\affil[1]{Mediterranean Institute of Technology, South Mediterranean University, Tunis, Tunisia}
\affil[2]{Dracodes, Tunis, Tunisia}
\affil[3]{National School of Computer Science, University of Manouba, Manouba, Tunisia}

\renewcommand{\headeright}{Preprint}
\renewcommand{\undertitle}{Preprint}
\renewcommand{\shorttitle}{LLM-Agent-UMF (A. Ben Hassouna et al.)}

\hypersetup{
pdftitle={LLM-Agent-UMF: LLM-based Agent Unified Modeling Framework for Seamless Design of Multi Active/Passive Core-Agent Architectures},
pdfsubject={cs.SE, cs.AI, cs.CL, cs.MA, cs.CR},
pdfauthor={Amine Ben Hassouna, Hana Chaari, Ines Bel-Haj},
pdfkeywords={LLM-based agent, Software architecture, Modularity, Security, Privacy, Safety, Core-agent classification, Multi-core agent},
}

\begin{document}
\maketitle

\def\thefootnote{§}\footnotetext{Equal contribution.}\def\thefootnote{\arabic{footnote}}

\begin{center}
\fbox{%
\begin{minipage}{0.7\textwidth}
This preprint is the manuscript submitted to and accepted in \href{https://www.sciencedirect.com/journal/information-fusion}{\textit{Information Fusion}}. The published article is available at \href{https://doi.org/10.1016/j.inffus.2025.103865}{https://doi.org/10.1016/j.inffus.2025.103865}. This manuscript is available under the \href{https://creativecommons.org/licenses/by-nc-nd/4.0/}{\textit{CC BY-NC-ND} license}.
\end{minipage}}
\end{center}

\vspace{0.5cm}

\begin{abstract}
In an era where vast amounts of data are collected and processed from diverse sources, there is a growing demand for sophisticated AI systems capable of intelligently fusing and analyzing this information. To address these challenges, researchers have turned towards integrating tools into LLM-powered agents to enhance the overall information fusion process. However, the conjunction of these technologies and the proposed enhancements in several state-of-the-art works followed a non-unified software architecture, resulting in a lack of modularity and terminological inconsistencies among researchers. To address these issues, we propose a novel LLM-based Agent Unified Modeling Framework (LLM-Agent-UMF) that establishes a clear foundation for agent development from both functional and software architectural perspectives, developed and evaluated using the Architecture Tradeoff and Risk Analysis Framework (ATRAF). Our framework clearly distinguishes between the different components of an LLM-based agent, setting LLMs and tools apart from a new element, the core-agent, which plays the role of central coordinator. This pivotal entity comprises five modules: planning, memory, profile, action, and security---the latter often neglected in previous works. By classifying core-agents into passive and active types based on their authoritative natures, we propose various multi-core agent architectures that combine unique characteristics of distinctive agents to tackle complex tasks more efficiently. We evaluate our framework by applying it to thirteen state-of-the-art agents, thereby demonstrating its alignment with their functionalities and clarifying overlooked architectural aspects. Moreover, we thoroughly assess five architecture variants of our framework by designing new agent architectures that combine characteristics of state-of-the-art agents to address specific goals. Throughout this evaluation, we leveraged ATRAF's Architectural Framework Tradeoff and Risk Analysis Method (AFTRAM), identifying quality attribute goals, developing scenarios, and analyzing architectural risks, which provided clear insights into potential improvements and highlighted challenges involved in both designing new agents and combining existing ones.
\end{abstract}


\section{Introduction}
\label{sec:introduction}

Large Language Models (LLMs) excel in tasks like language modeling, question answering, sentiment analysis, Natural Language Understanding (NLU), commonsense reasoning and knowledge fusion {\cite{67,163}}. However, standalone LLMs lack other skills such as information retrieval, mathematical reasoning, code evaluation, and numerous others. These functional shortcomings can be managed by AI agents leveraging external tools, knowledge repositories, and human feedback. An autonomous agent is a system interacting with an environment, sensing it, and acting on it over time following a certain agenda {\cite{136}}. While there are different classes of agents, in this paper we will be focusing on LLM-based agents which, by combining the capabilities of LLMs and autonomous agents, can achieve a broad range of tasks~{\cite{67}}.

In fact, the shift towards more natural conversational interfaces powered by LLMs is transforming the way humans engage with agents, enabling seamless and intuitive interactions. Furthermore, LLM-based agents play an essential role as information fusion intermediaries by intelligently synthesizing data from heterogeneous sources and providing meaningful insights to human users or other AI systems. These LLM-powered agents have now become dominant in the landscape of AI agents, and they are regarded as potential steppingstones towards Artificial General Intelligence (AGI), offering hope for the development of AI agents that can adapt to diverse scenarios {\cite{135}}.

Understanding the potential of these agents and improving them necessitates a deep understanding of their structure. In LLM-based agents, besides the LLM that handles reasoning, there are other components that oversee the execution of tasks, ensure the security of the agent, and handle its memory {\cite{135}}. However, merely pinpointing existing functionalities is insufficient for developers and researchers. To provide a comprehensive overview of these agents, the survey {\cite{wang2024}} proposed a comprehensive framework for their construction, consisting of four main modules which will be discussed thoroughly in the next section. Although it provides valuable insights into the functionality of each module in an agent, it overlooks their delineation from a software architectural perspective. This perspective is crucial for developers to establish a common base architecture to build upon and for researchers to improve. Our analysis of the state-of-the-art LLM-based agents reveals common limitations: the complexity of implementation, resulting in unstructured and ambiguous software architecture; lack of modularity and composability, making components non-reusable by other agent-based solutions; and difficulty in maintainability and introducing improvements to existing agents.

In response to these limitations, we propose the LLM-based Agent Unified Modeling Framework (LLM-Agent-UMF). To the best of our knowledge, our framework is the first to emphasize a clear delineation of each component within an LLM-powered agent and define their interactions within specified boundaries from both architectural and functional perspectives. In addition to traditional components like LLMs and tools, we introduce a new unit within the agent, which we label as the ``core-agent''. This pioneering entity is further classified into two types---active and passive core-agents---which enhance our understanding of each component's capabilities and accurately describe the dynamics between modules within the agent. As a result, we alleviate the complexity of the architecture and improve the reusability of the components, promoting a shift from multi-agent systems to multi-core agents.

Throughout our paper, we highlight several advantages offered by the LLM-Agent-UMF, from resolving terminological ambiguities to the introduction of enhancing modules like the security module. To validate the reliability of our framework, we apply it to existing solutions and identify the modules within each agent. This approach enables us to assess the feasibility and requirements for merging one agent with another, ultimately aiming to create an agent with fully enhanced capabilities.

\vspace{0.1cm}
\noindent Compared with previous works, the five main contributions of this paper can be summarized as follows:
\begin{enumerate}
    \item We introduce a new terminology, core-agent, as a structural sub-component in LLM-based agents to improve modularity and promote more effective and precise communication among researchers and contributors in the field of agents and LLM technologies.

    \item We model the internal structure of a core-agent by adapting the framework suggested by {\cite{wang2024}} that was originally meant to describe the whole agent from an abstract functional perspective.

    \item We improve our modeling framework by augmenting the core-agent with a security module and introducing new methods within other modules.

    \item We classify core-agents into active and passive ones, explaining their differences and similarities, and highlighting their unique advantages.

    \item Finally, we introduce various multi-core agent architectures, emphasizing that the hybrid one-active-many-passive core-agent architecture is the optimal setup for LLM-based agents.
\end{enumerate}

The rest of this paper is organized as follows. Section~\ref{sec:related-work} provides a background on LLM-based agents and reviews relevant state-of-the-art works. Section~\ref{sec:methodology} outlines the methodology employed in developing the LLM-Agent-UMF, including the literature review process, quality requirements, design principles, and validation approach. Section~\ref{sec:llm-agent-umf} introduces our LLM-based Agent Unified Modeling Framework, detailing its components and proposed multi-core agent architectures. Section~\ref{sec:evaluation} evaluates the efficiency of our framework by applying it to state-of-the-art agents and designing novel multi-core agent architectures. Finally, Section~\ref{sec:conclusion} summarizes key findings and discusses future challenges and directions for enhancing the framework.

\section{Related Work}
\label{sec:related-work}

We will start off this section by providing a comprehensive overview of the key concepts that serve as the fundamental basis for our work. First, tool-augmented LLMs are a major advancement in Natural Language Processing (NLP) that combine the language understanding and generation capabilities of LLMs with the ability to interface with external tools and Application Programming Interfaces (APIs). For instance, TALM {\cite{126}} introduces models which can leverage a wide range of functionalities, from information retrieval to task planning and execution.

By incorporating tool-augmented LLMs, LLM-based autonomous agents exhibit exceptional proficiency in NLP tasks {\cite{cheng2024}}, including reasoning {\cite{23}}, programming {\cite{90}}, and text generation, surpassing other types of agents in these areas. Moreover, they address several limitations of standalone LLMs, such as context length constraints and the inability to utilize tools. This development marks a significant breakthrough in the scientific community, explaining the recent surge in the adoption of LLM-based agents.

Researchers and practitioners across various disciplines are leveraging these agents to tackle complex problems and drive innovation in fields such as gaming {\cite{58}} and other professional domains that require specialized expertise {\cite{55}}. Additionally, the trend towards integrating LLMs in scientific research, namely in chemistry {\cite{132}}, highlights their transformative potential, promising to drive forward new discoveries and applications.

Upon examining existing agents, we detected a notable absence of direct security measures or guardrails within the agents to ensure the protection of sensitive information and enhance overall system integrity. Indeed, the level of autonomy in LLM-based agents raises significant concerns regarding ethical use, malicious data, privacy and robustness {\cite{111}}. Notably, one critical risk involves jailbreaks which can be mitigated through the implementation of more robust monitoring and control mechanisms to detect and respond to jailbreaks during deployment {\cite{139}}. Moreover, data privacy remains a persistent challenge in any software system, including agents. To address this, various methods have been developed to safeguard against data extraction attempts from prompts, such as leveraging privacy-preserving algorithms for prompt learning {\cite{138}}. These findings prompted us to place an emphasis on this often-neglected aspect and include security measures in our framework.

\begin{figure}[h]
    \centering
    \includegraphics[width=0.8\linewidth]{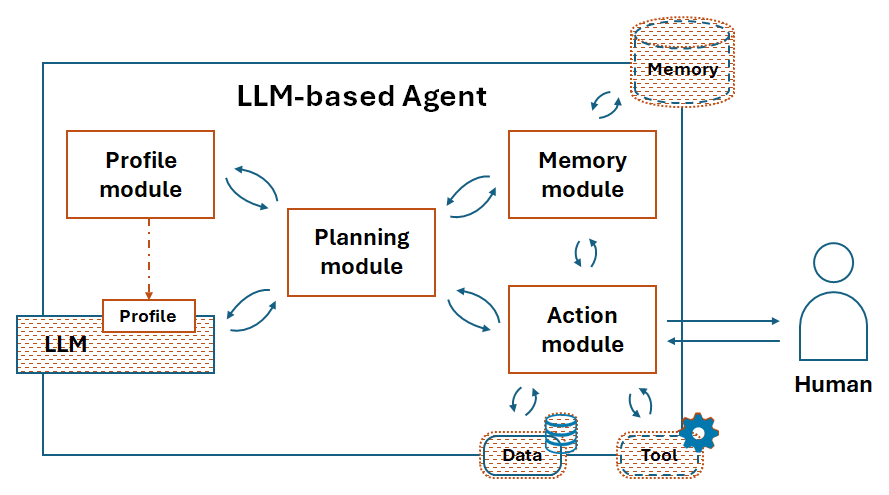}
    % Comment: This figure illustrates the framework for LLM-based agents proposed by survey (Wang, 2024), depicting four key modules: Profile, Memory, Planning, and Action. It highlights the functional structure of an LLM-based agent but lacks clear delineation of components like the LLM, tools, and data sources, leading to potential ambiguities in software architecture.
    \caption{Framework proposed by survey {\cite{wang2024}} for LLM-based agents}
    \label{fig:framework-proposed-in-first-survey}
\end{figure}

Besides security assurance, the development of AI systems, particularly agents, must adhere to fundamental software development principles such as modularity, composability, and maintainability. These principles enhance the flexibility, scalability, and adaptability of AI systems, enabling them to effectively meet the evolving needs of both users and businesses.

Existing agents do not consistently adhere to these principles as they do not focus primarily on architectural design. This underscores the importance of architectural frameworks as blueprints for LLM-based agents, highlighting their essential role in releasing the full potential of these systems. Such frameworks enable the development of modular, robust, extensible, and interoperable designs. They provide the necessary scaffolding to build increasingly capable and reliable agents capable of tackling complex real-world problems.

For example, the paper {\cite{cheng2024}} exploring LLM-based intelligent agents points out 5 main axes: Planning, Memory, Rethinking, Environment, and Action. Despite their attempt to delineate between LLMs, environment and tools, they did not define the software components of the agent. Likewise, the framework proposed by the survey {\cite{wang2024}} identifies the structure and applications of LLM-powered autonomous agents and highlights four key modules as shown in Figure~\ref{fig:framework-proposed-in-first-survey}: the Profile module, the Memory module, the Planning module, and the Action module.

The profile module serves to delineate the diverse roles of the LLM. Meanwhile, the memory module retains internal logs encompassing the agent's past thoughts, actions, and observations within its dynamic environment, including interactions with users. The planning module guides the agent in decomposing overarching tasks into manageable steps or subtasks, enhancing responsiveness by leveraging past behaviors in future plans. Together, these modules significantly influence the action module, which translates the agent's decisions into specific outputs harnessing external tools to extend the agent's capabilities {\cite{wang2024}}.

While this framework effectively addresses the functionalities of the agent, it does present some areas for improvement. First, there are functional overlaps in certain modules, particularly the overlap of reflective activities between planning and memory modules. Second, the definition of memory is ambiguous as will be discussed in Section~\ref{sec:comparison-prior-frameworks}, leading to confusion as the term is used to represent different concepts without clear distinction. Lastly, as illustrated in Figure~\ref{fig:framework-proposed-in-first-survey}, it is not explicitly stated whether the LLM, tools, data sources, and memory are part of the agent. This fuzzy distinction between the functionalities of each module foster division between software developers and leads to incompatibility and discourages reusability. In the next section, we introduce the main components of our framework, explain the rationale behind their inclusion, and highlight how they solve the limitations present in other works.

\section{Methodology}
\label{sec:methodology}

The Software Development Life Cycle (SDLC) provides a systematic framework for developing robust and scalable systems, encompassing six key phases: Requirements Specification, Software Design, Implementation, Testing, Deployment, and Maintenance. These phases ensure that complex systems, such as LLM-based agents, are designed and delivered with clarity, efficiency, and adaptability. LLM-Agent-UMF focuses primarily on the Requirements Specification and Software Design phases, as these are critical for establishing precise functional and non-functional requirements, crafting architectural blueprints that balance performance, flexibility, and maintainability.

Thoughtful Requirements Specification informs the Software Design phase by defining quality attributes and design principles, which streamline subsequent phases. Modular designs enable efficient Implementation by allowing independent component development, reduce Testing efforts by addressing issues early, facilitate scalable Deployment through strategic architectural choices, and support maintainable architectures for long-term Maintenance. However, to ensure the robustness of the architectural design, a structured evaluation method is essential.

The Architecture Tradeoff and Risk Analysis Framework (ATRAF)~\cite{benhassouna2025atraf} provides a robust evaluation method through its Architectural Framework Tradeoff and Risk Analysis Method (AFTRAM), tailored for Architectural Frameworks like the LLM-Agent-UMF, which qualifies as such due to its process-driven methodology, lifecycle support, and cross-domain applicability. AFTRAM’s iterative four-phase spiral process was employed to develop and evaluate LLM-Agent-UMF, with each phase mapped to IMRaD sections per the ATRAF-driven IMRaD Methodology~\cite{benhassouna2025atrafimrad} for transparency, rigor, and reproducibility: Phase I (Multi-Context Scenario and Requirements Gathering) elicits scenarios and quality requirements, spanning the Introduction and Methodology sections; Phase II (Architectural Framework Views and Scenario Realization) presents architectural views, detailed in the Methodology section; Phase III (Attribute-Specific Analyses) evaluates quality attributes, covered in the Evaluation and Discussion section; and Phase IV (Sensitivity, Tradeoff, and Risk Analysis) identifies tradeoffs and risks, also in the Evaluation and Discussion section, with interpretations in its Discussion component, presenting a final snapshot of the AFTRAM process.

\subsection{Literature Review Process}
\label{sec:literature-review-process}

To develop the LLM-based Agent Unified Modeling Framework (LLM-Agent-UMF), a comprehensive literature review was conducted to identify gaps in existing LLM-based agent frameworks and inform a modular, secure, and scalable design. The review systematically explored scholarly databases and digital libraries, including ScienceDirect, Scopus, Springer, Nature, IEEE Xplore, ACM Digital Library, ACL Anthology, NeurIPS proceedings, AAAI library, arXiv, Google Scholar, and Semantic Scholar. The process targeted literature on tool-augmented LLMs, LLM-based agents, planning strategies, information fusion, AI ethics, privacy, security, software modeling best practices, and terminological consistency in AI systems, ensuring a robust foundation for addressing architectural and functional deficiencies. The review was guided by thirteen specific objectives, each addressing a critical gap or requirement that shaped the framework's design.

\begin{enumerate}
    \item \textbf{Modularity Inspiration from Biological Models}: To ensure a modular structure, the review explored cognitive science literature, focusing on human brain modularity, where distinct regions coordinate for cognitive tasks \cite{131}. This informed the application of the Single Responsibility Principle (SRP) to delineate the core-agent's modules (planning, memory, profile, action, security), ensuring each handles specific functions independently.

    \item \textbf{Structured Analysis for Planning Module}: To address the lack of clear component delineation in prior frameworks, the review analyzed LLM-based agent planning literature \cite{108}. This led to a novel four-aspect analysis (process, strategies, techniques, feedback sources) for the planning module, providing a structured approach to task decomposition and plan generation.

    \item \textbf{Human-inspired Task Decomposition}: Drawing from cognitive psychology, the review examined human cognitive strategies for breaking down complex tasks \cite{131}. This informed the design of iterative and non-iterative decomposition approaches in the planning module, enhancing its ability to handle complex problems systematically.

    \item \textbf{Criteria-based Technique Selection}: To develop a methodology for selecting planning techniques, the review synthesized literature on planning algorithms \cite{108,149}. Criteria such as task complexity and contextual comprehension were derived, enabling the planning module to choose between rule-based and language model-powered techniques effectively.

    \item \textbf{System Boundary Delineation}: To ensure interoperability with external tools and humans, the review studied system architecture literature \cite{113}. This established clear system boundaries for the planning module's feedback sources (human, tool, sibling core-agent), promoting seamless integration within the agent system.

    \item \textbf{Addressing Terminology Gaps in Memory Module}: To resolve unconventional memory terminology in prior works, the review aligned memory scope with human memory models from cognitive science \cite{131}. Terms like short-term and long-term memory were adopted, replacing ambiguous Unified and Hybrid Memory terms, enhancing clarity and consistency.

    \item \textbf{Architectural Relevance for Memory Location}: Prioritizing software architecture relevance, the review analyzed software engineering literature to focus on memory location (embedded vs. extension) \cite{104}. This approach addressed gaps in prior frameworks that emphasized scope over architectural considerations, improving memory management design.

    \item \textbf{Categorization of Memory Formats}: To clarify memory representation, the review synthesized literature on data representation formats (natural language, embeddings, SQL databases, structured lists) \cite{133,127}. This led to the novel three-perspective analysis (scope, location, format) for the memory module, addressing literature gaps in representation clarity.

    \item \textbf{Best Practice Alignment for Security Module}: To ensure the security module adhered to modern standards, the review examined industry best practices and AI security standards \cite{eu_ai_act,iso_22989_2022}. The Secure by Design principle was adopted, integrating security from the outset to address high-risk AI deployment challenges.

    \item \textbf{Advancement Over C.I.A. Triad Usage}: To overcome limitations in prior works using the C.I.A. triad (Confidentiality, Integrity, Availability) as a risk-classification schema, the review analyzed cybersecurity literature \cite{151}. The triad was integrated as a foundational architectural principle, guiding the security module's design for robust protection.

    \item \textbf{Comprehensive Security Research}: To address multifaceted security challenges, the review explored AI security literature covering asset protection, threat defense, incident mitigation, and privacy compliance \cite{111,112}. This informed the security module's layered safeguarding mechanisms (prompt, response, data privacy safeguarding).

    \item \textbf{Guardrail Methodology Analysis}: To operationalize security measures, the review analyzed guardrail methodologies from LLM service providers and open-source communities \cite{112,144}. Rule-based and LLM-based guardrails were identified, forming the security module's operational layer for effective threat mitigation.

    \item \textbf{Regulatory Compliance for Trust Boundaries}: To ensure compliance with data protection obligations, the review examined data protection regulations \cite{eu_ai_act}. Transparent trust boundaries were architected for the security module, ensuring privacy and user confidence in interactions with external tools.

\end{enumerate}

The literature review process was iterative, with findings cross-referenced to validate gaps and ensure alignment with software engineering and AI development best practices. By addressing modularity, terminology, security, and architectural clarity, the review provided a solid foundation for the LLM-Agent-UMF, enabling it to overcome the limitations of prior frameworks and establish a unified, extensible, and secure approach to LLM-based agent design.

\subsection{Quality Requirements}
\label{sec:quality-requirements}

The LLM-Agent-UMF was designed to meet a comprehensive set of non-functional quality requirements to ensure robust, efficient, and trustworthy LLM-based agent systems. These requirements, derived from an analysis of existing frameworks and software engineering best practices, address modularity, security, scalability, and other critical attributes that enhance the framework's applicability and the performance of agents designed using it. The following quality requirements were explicitly addressed in the framework's design, while others such as recovery were considered for future research.

\begin{itemize}
    \item \textbf{Flexibility}: Allows the framework to adapt to varying requirements, environments, or use cases, ensuring agents can handle diverse tasks and dynamic contexts effectively.
    \item \textbf{Extensibility}: Enables the framework to be expanded with new features or capabilities, supporting the integration of additional modules or functionalities without disrupting existing components.
    \item \textbf{Reusability}: Facilitates the use of framework components across different agents or applications, reducing duplication and enhancing development efficiency.
    \item \textbf{Modifiability}: Enables straightforward alterations or updates to accommodate changes, maintaining system relevance and functionality over time.
    \item \textbf{Maintainability}: Ensures that the framework and its agents can be easily updated or fixed, supporting long-term system evolution and operational continuity.
    \item \textbf{Scalability}: Supports handling increased loads or growth, allowing agents to manage expanding tasks or user demands without performance degradation.
    \item \textbf{Dynamic Adaptability}: Allows agents to adjust to changing conditions or requirements at runtime, improving responsiveness and versatility.
    \item \textbf{Resilience and Fault Tolerance}: Ensures agents can withstand disruptions and continue operating despite failures or errors, enhancing reliability and robustness in unpredictable environments.
    \item \textbf{Performance}: Optimizes operational efficiency, including speed, responsiveness, and resource usage, to deliver high-quality user experiences.
    \item \textbf{Availability}: Ensures reliable system operation and accessibility, minimizing downtime and maintaining service reliability for users.
    \item \textbf{Interoperability}: Facilitates seamless interaction with other systems or components, enabling effective information exchange and integration.
    \item \textbf{Security}: Protects against unauthorized access, threats, or data breaches, incorporating integrity to maintain the accuracy, completeness, and uncorrupted state of data, processes, and components, thereby safeguarding system reliability and user confidence.
    \item \textbf{Privacy}: Protects personal or sensitive data from unauthorized access or disclosure, ensuring compliance with data protection standards and fostering user trust.
    \item \textbf{Safety and Ethical Alignment}: Ensures adherence to ethical standards and principles, promoting user trust and societal acceptance of agent actions.
\end{itemize}

\subsection{Design Principles}
\label{sec:design-principles}

The LLM-Agent-UMF is grounded in a set of software engineering design principles that guide its development and serve as recommendations for architects and designers creating LLM-based agents based on the framework. These principles, derived from software engineering best practices and tailored to address the challenges of LLM-based agent systems, ensure modularity, security, scalability, and clarity. They support the quality requirements outlined in Section~\ref{sec:quality-requirements}, such as \textit{Flexibility}, \textit{Maintainability}, and \textit{Security}, by providing actionable strategies for architectural design. The principles are organized into three categories---general architectural principles, security and privacy principles, and framework-level design principles---each contributing to the framework's robustness and applicability.

\vspace{0.5cm}
\noindent\textbf{General architectural principles:}
\begin{itemize}
    \item \textbf{Terminological Consistency}: Uniform use of terms and concepts ensures \textit{Maintainability} and \textit{Modifiability} by reducing ambiguity. The lack of standardized terminology across researchers and practitioners has led to discrepancies in describing LLM-based agent components, making this a primary concern for clear communication and collaboration \cite{153}.
    \item \textbf{Modularity, Single Responsibility Principle (SRP), and Separation of Concerns}: Systems are broken into smaller, independent modules, each with a single responsibility and distinct functionality, to enhance \textit{Maintainability}, \textit{Scalability}, \textit{Reusability}, \textit{Modifiability}, and \textit{Performance}. This combined principle ensures clear delineation of responsibilities, reducing functional overlap \cite{113}.
    \item \textbf{Open-Closed Principle (OCP) and Loose Coupling}: Systems are open for extension but closed for modification, with components designed to have minimal dependencies, supporting \textit{Modifiability}, \textit{Maintainability}, \textit{Scalability}, \textit{Flexibility}, and \textit{Extensibility}. This principle enables new functionalities without altering existing components and ensures independent updates \cite{160}.
    \item \textbf{Composability}: Components are designed to be combined or reused to build larger systems, supporting \textit{Flexibility}, \textit{Scalability}, \textit{Maintainability}, \textit{Reusability}, and \textit{Extensibility}, enabling versatile system configurations.
\end{itemize}

\noindent\textbf{Security and privacy principles:}
\begin{itemize}
    \item \textbf{Security by Design (SbD)}: Security is integrated into the development process from the outset to achieve \textit{Security} and \textit{Integrity}, ensuring robust protection against threats \cite{151, giorgini2005}.
    \item \textbf{Privacy by Design (PbD)}: Privacy considerations are integrated from the start to achieve \textit{Privacy}, protecting sensitive data and aligning with data protection standards~\cite{delreal2024sss, obiokafor2025integrating}.
    \item \textbf{C.I.A. Triad}: The \textit{Confidentiality}, \textit{Integrity}, and \textit{Availability} (C.I.A.) triad serves as a foundational principle to ensure secure and reliable systems, safeguarding data confidentiality, maintaining data and process integrity, and ensuring system availability \cite{151}.
    \item \textbf{Trust Boundaries Definition}: Zones of trust are clearly defined to manage security and access control, supporting \textit{Security}, \textit{Integrity}, and \textit{Privacy}, ensuring compliance with data protection requirements~\cite{112}.
    \item \textbf{Defense in Depth}: Multiple layers of security controls are implemented to protect against threats, achieving \textit{Security} and \textit{Integrity}, enhancing system resilience \cite{175}.
\end{itemize}

\noindent\textbf{Framework-level design principles:}
\begin{itemize}
    \item \textbf{Technology-Agnostic Design}: The framework avoids dependency on specific technologies to promote \textit{Flexibility}, \textit{Maintainability}, and \textit{Interoperability}, allowing architects and designers to select technologies best suited for their LLM-based agents while ensuring the framework's flexibility to support these choices.
    \item \textbf{Variability Management}: Variations in requirements or configurations are handled systematically, supporting \textit{Flexibility}, \textit{Extensibility}, \textit{Maintainability}, and \textit{Reusability}, accommodating diverse agent \nobreak{designs}.
    \item \textbf{Extensibility Mechanisms}: Extension points, such as pluggable modules and support for multiple LLMs, enable future enhancements without core changes, supporting \textit{Extensibility}, \textit{Maintainability}, \textit{Flexibility}, and \textit{Reusability}.
    \item \textbf{Constraint-based Design}: Rules and boundaries are established to guide the framework's design, ensuring consistent and predictable behavior, supporting \textit{Performance}, \textit{Maintainability}, \textit{Safety}, \textit{Security}, \textit{Integrity}, and \textit{Availability}.
\end{itemize}

These design principles collectively ensure that the LLM-Agent-UMF is robust, adaptable, and secure, while providing a clear blueprint for designing LLM-based agents. By adhering to these principles, architects can create agents that meet the quality requirements outlined in Section~\ref{sec:quality-requirements}, fostering \textit{Modularity}, \textit{Scalability}, and \textit{Trustworthiness} in complex AI systems.

\subsection{Contributions: Core-Agent as Keystone Component}
\label{sec:core-agent-contributions}

\begin{figure}[h]
    \centering
    \includegraphics[width=0.8\linewidth]{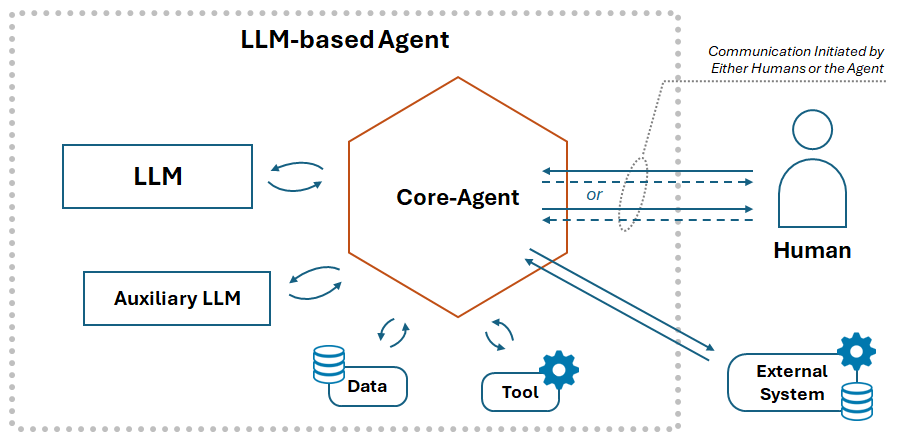}
    % Comment: This figure presents the core-agent as the central coordinating component within an LLM-based agent. It shows the core-agent interacting with the LLM, tools, and the environment, emphasizing its role as a controller that translates high-level goals into actions and facilitates communication, enhancing modularity and clarity in agent architecture.
    \caption{The core-agent as the central component of LLM-based agents}
    \label{fig:core-agent-as-the-central-component-of-llm-based-agent}
\end{figure}

Prior frameworks for LLM-based agents, such as those proposed by \cite{wang2024,cheng2024}, and \cite{34}, have primarily focused on the functional perspective, detailing capabilities like planning, memory, and action. However, these works often neglect issues from a software architectural perspective, failing to delineate the boundaries of different software components within the agent. This lack of architectural clarity results in ambiguous designs, reduced \textit{Modularity}, and challenges in \textit{Reusability} and \textit{Maintainability}. LLM-Agent-UMF addresses these limitations by clearly delineating software components, introducing the core-agent as the keystone component that serves as the central coordinator, as depicted in Figure~\ref{fig:core-agent-as-the-central-component-of-llm-based-agent}.

\begin{figure}[h]
    \centering
    \includegraphics[width=\linewidth]{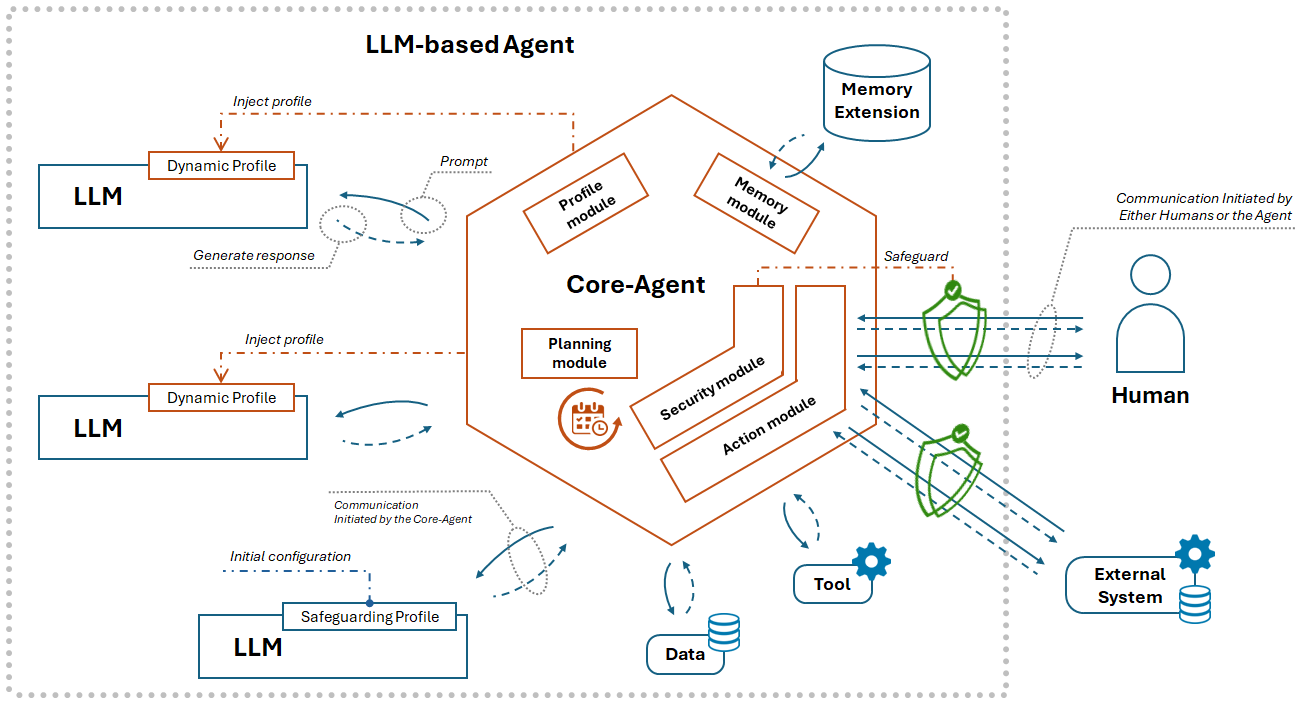}
    % Comment: This figure provides an overview of the core-agent's internal structure within an LLM-based agent, depicting five modules: Planning, Memory, Profile, Action, and Security. It illustrates how these modules collaborate within the core-agent, interacting with the LLM and tools, to ensure modularity, extendibility, and maintainability, inspired by the human brain's modular design.
    \caption{Overview of the core-agent internal structure within an LLM-based agent}
    \label{fig:overview-of-core-agent-internal-structure-in-llm-based-agent}
\end{figure}

The core-agent is not a newly invented component but a label to denote an existing functional element that has been previously implicit or unnamed in past frameworks or architectures. It acts as the pivotal interface, orchestrating interactions among Large Language Models (LLMs), tools, and the environment to translate high-level goals into actionable outcomes, enhancing clarity and reusability \cite{153}. By providing a distinct, modular entity, the core-agent mitigates the complexity of agent architectures, particularly in multi-LLM setups, serving as the primary communication hub. Its internal structure, comprising five modules---planning, memory, profile, action, and security---operationalizes this role, as detailed in Section~\ref{sec:llm-agent-umf} and illustrated in Figure~\ref{fig:overview-of-core-agent-internal-structure-in-llm-based-agent}. This section outlines the framework's key contributions, mapping each to the design principles from Section~\ref{sec:design-principles} that enable actionable concepts to ensure the quality requirements from Section~\ref{sec:quality-requirements}, such as \textit{Modularity}, \textit{Security}, and \textit{Scalability}.

\vspace{0.2cm}
\noindent \textbf{The framework's contributions are as follows:}
\begin{itemize}
    \item \textbf{Novel Core-Agent Terminology}: Labeling the central coordinating component as the ``core-agent'' addresses terminological discrepancies in LLM-based agent research, fostering consistent communication. This contribution is driven by \textit{Terminological Consistency} to resolve ambiguities, \textit{Modularity, SRP, and Separation of Concerns} to delineate the core-agent as a distinct entity, and \textit{Constraint-based Design} to establish a standardized term for consistent framework behavior, supporting \textit{Maintainability} and \textit{Modifiability} \cite{153}.

    \item \textbf{Dedicated Security Module}: Introducing a security module ensures ethical and secure operations through layered safeguarding mechanisms, addressing critical gaps in prior frameworks that neglected security. This contribution is guided by \textit{Security by Design}, which embeds robust protection from the outset, \textit{Privacy by Design}, which prioritizes sensitive data protection, \textit{C.I.A. Triad}, which ensures \textit{Confidentiality}, \textit{Integrity}, and \textit{Availability}, \textit{Trust Boundaries Definition}, which secures external interactions, and \textit{Defense in Depth}, which provides layered \textit{Resilience}. Additionally, \textit{Modularity, SRP, and Separation of Concerns} isolate security functions, and \textit{Constraint-based Design} enforces security constraints, enhancing \textit{Trustworthiness}, \textit{Resilience}, and \textit{Privacy} \cite{111,112,151}.

    \item \textbf{Enhanced Module Delineation}: Refining the core-agent's modules with new methods and perspectives (e.g., four-aspect planning analysis, three-perspective memory analysis) improves \textit{Flexibility}, \textit{Interoperability}, and \textit{Performance}. This contribution is enabled by \textit{Modularity, SRP, and Separation of Concerns} for clear module responsibilities, \textit{Open-Closed Principle and Loose Coupling} for independent \textit{Extensibility}, \textit{Variability Management} for diverse module methods, \textit{Extensibility Mechanisms} for adding new methods, and \textit{Constraint-based Design} for consistent module behavior, as detailed in Section~\ref{sec:llm-agent-umf}.

    \item \textbf{Active and Passive Core-Agent Classification}: Classifying core-agents as active (authoritative, encompassing all five modules for complex task management) or passive (non-authoritative, limited to action and security modules for simplified execution) provides a novel taxonomy to address architectural ambiguities in prior agent designs. This classification enhances role clarity, enabling active core-agents to handle planning and decision-making while passive core-agents streamline task execution with minimal synchronization needs. It is driven by \textit{Modularity, SRP, and Separation of Concerns} for distinct role delineation, \textit{Composability} for combining core-agents in multi-core setups, \textit{Variability Management} for flexible role assignments, and \textit{Constraint-based Design} for predictable interactions, supporting \textit{Scalability} and \textit{Modularity} \cite{153}.

    \item \textbf{Multi-Core Agent Architectures}: Proposing uniform (multi-passive or multi-active) and hybrid (one-active-many-passive and many-active-many-passive) multi-core architectures addresses \textit{Scalability} and \textit{Interoperability}, challenges of multi-agent systems. The one-active-many-passive architecture is optimal, balancing the simplicity of passive core-agents with the managerial capabilities of an active core-agent to dynamically allocate resources and handle diverse tasks. This contribution is enabled by \textit{Composability} for seamless core-agent integration, \textit{Open-Closed Principle and Loose Coupling} for extensible and scalable designs, \textit{Technology-Agnostic Design} for supporting diverse LLMs and tools, \textit{Variability Management} for flexible configurations, \textit{Extensibility Mechanisms} for accommodating system growth, and \textit{Constraint-based Design} for consistent \textit{Performance}, ensuring \textit{Scalability}, \textit{Flexibility}, and robustness \cite{40}.
\end{itemize}

\subsection{Comparison with Prior Frameworks}
\label{sec:comparison-prior-frameworks}

LLM-Agent-UMF introduces a modular architecture centered on the core-agent, comprising five internal modules: planning, memory, profile, action, and security, designed with a technology-agnostic approach and principles like \textit{Modularity}, \textit{Single Responsibility Principle (SRP)}, \textit{Separation of Concerns}, \textit{Open-Closed Principle}, \textit{Loose Coupling}, and \textit{Constraint-based Design} to ensure \textit{Flexibility}, \textit{Extensibility}, and \textit{Maintainability} across all modules. This section compares our framework primarily with the prior framework proposed by~\cite{wang2024}, which defines LLM-based agents from a functional perspective through four modules: planning, profile, memory, and action. We occasionally reference differences with \cite{cheng2024} to highlight additional distinctions, without focusing on it. Our framework addresses limitations in~\cite{wang2024}, such as functional overlaps, ambiguous terminology, and lack of architectural delineation, as illustrated in \cite{wang2024}'s Figure~\ref{fig:framework-proposed-in-first-survey}, which does not clarify component boundaries, and introduces novelties, such as the security module, to ensure modularity, clarity, and trustworthiness.

\begin{itemize}
    \item \textbf{Planning Module}: In \cite{wang2024}, the planning module guides the agent in decomposing tasks into subtasks, leveraging past behaviors for responsiveness. Our framework aligns with this functional perspective but repositions the planning module within the core-agent, enabling collaboration with sibling modules, such as memory for memory-augmented planning \cite{108}. We introduce a novel four-aspect analysis (process, strategies, techniques, feedback sources) to model the aspects of planning from multiple perspectives, as opposed to \cite{wang2024}'s abstract functional description. This structured approach mitigates functional overlaps, particularly with memory, and clarifies planning responsibilities, supporting \textit{Modularity} and \textit{Composability}. Similarly, \cite{cheng2024}'s ``Rethinking'' module, functionally akin to planning, guides future actions through introspection on past actions and feedback, but its separation as a distinct module and lack of a core-agent structure limit architectural clarity compared to our integrated approach. This module's design was further guided by \textit{Extensibility Mechanisms} to support flexible planning strategies and techniques.

    \item \textbf{Memory Module}: \cite{wang2024} defines the memory module as retaining internal logs of thoughts, actions, and observations, using ambiguous terms like Unified Memory (short-term, in-context learning) and Hybrid Memory (short- and long-term). This abstract approach leads to functional overlaps, particularly with reflection on the memory, which \cite{wang2024} includes in the memory module. Our framework redefines memory through a three-perspective analysis (scope, location, format) to manage storage and retrieval of task-related information, building on \cite{cheng2024}'s short-term and long-term categorization but enhancing it with alignment to established human memory models \cite{131} and moving all reflection aspects to the planning module. This adaptation eliminates overlaps, resolves ambiguous terminology, and enhances architectural clarity, ensuring \textit{Maintainability}. Similarly, \cite{cheng2024} treats memory as a state repository, with reflection handled by its ``Rethinking'' module, functionally akin to planning, aligning with our functional approach, but its lack of a core-agent structure and structured analysis reduces clarity compared to our modular design.

    \item \textbf{Profile Module}: \cite{wang2024}'s profile module delineates the LLM's diverse roles, supporting three methods: Handcrafted In-Context Learning, LLM-Generation, and Dataset Alignment. Our framework builds on \cite{wang2024}'s role-defining function, refining the conceptual and functional clarity of its three methods (Handcrafted In-Context Learning, LLM-Generation, Dataset Alignment) to support \textit{Dynamic Adaptation}, and introduces a novel fourth method, Fine-tuned Pluggable Modules, which enhances \textit{Performance} by reducing context size and memory footprint for efficient inference. These enhanced and new methods, integrated within the core-agent, ensure clear separation from the LLM, addressing \cite{wang2024}'s lack of architectural boundaries and improving \textit{Flexibility} and \textit{Performance}. Unlike \cite{wang2024}, \cite{cheng2024} does not define a distinct profile module, relying on LLM configuration within its quintuple model, which limits role-specific adaptability compared to our framework. This module was further shaped by \textit{Variability Management} and \textit{Extensibility Mechanisms} to enable diverse and pluggable role-defining methods.

    \item \textbf{Action Module}: In \cite{wang2024}, the action module translates decisions into outputs using external tools, influenced by planning, memory, and profile modules. Our framework retains this functional role, translating plans into executable actions guided by specific triggers and goals, but refines it with a four-aspect analysis (goal, trigger, action space, impact), introducing new triggers (Plan Following and API Call Request) to enhance action initiation clarity. Unlike \cite{wang2024}, which risks overlap between action and memory functions (e.g., Action via Memory Recollection), our framework separates these responsibilities, ensuring the action module focuses on execution while memory handles storage, thus improving \textit{Modularity} and \textit{Efficiency}. Similarly, \cite{cheng2024} defines action as tool usage or messaging, with its ``Rethinking'' module, functionally akin to planning, guiding future actions through introspection on past actions and feedback, whereas our framework integrates reflection within the planning module's feedback sources and action module's impact analysis for clearer architectural delineation within the core-agent. The action module was further guided by \textit{Extensibility Mechanisms} to support new triggers and flexible tool integration.

    \item \textbf{Security Module}: \cite{wang2024} does not include a security module, leaving gaps in addressing ethical use, privacy, and robustness, critical concerns for LLM-based agents \cite{111}. Our framework introduces a novel security module within the core-agent, ensuring ethical and secure operations through layered safeguarding mechanisms (prompt, response, data privacy). Likewise, \cite{cheng2024} omits a dedicated security module, focusing on cognitive and functional aspects without addressing architectural security, further underscoring our framework's novelty. This module's design was driven by \textit{Security by Design}, \textit{Privacy by Design}, \textit{C.I.A. Triad}, \textit{Trust Boundaries Definition}, and \textit{Defense in Depth} to ensure robust protection against ethical and security risks.
\end{itemize}

By adapting \cite{wang2024}'s modules to operate under the core-agent, introducing a security module, and addressing deficiencies also present in \cite{cheng2024}, the LLM-Agent-UMF overcomes functional overlaps, ambiguous terminology, and lack of architectural clarity.

\subsection{Validation Approach}
\label{sec:validation-approach}

The validation of the LLM-Agent-UMF was conducted through a scenario-driven approach, AFTRAM (Section~\ref{sec:methodology}), strategically designed to assess its capability to guide the architectural design of LLM-based agents without implementing them. The goal was to evaluate the framework's effectiveness in enhancing \textit{Modularity}, \textit{Scalability}, \textit{Reusability}, \textit{Maintainability}, \textit{Flexibility}, \textit{Security}, and \textit{Privacy}, as outlined in the quality requirements (Section~\ref{sec:quality-requirements}). This was achieved by applying the framework to thirteen state-of-the-art agents and designing five novel multi-core agent architectures, each targeting specific quality attributes. The agents were carefully selected to represent a diverse range of functionalities, including tool usage (Toolformer~\cite{22}, Confucius~\cite{2}, ToolAlpaca~\cite{19}, Gorilla~\cite{5}, ToolLLM~\cite{1}, GPT4Tools~\cite{21}), memory management (ChatDB~\cite{133}), planning (Chameleon~\cite{23}, LLM+P~\cite{155}), chemistry applications (ChemCrow~\cite{132}), security (LLMSafeGuard~\cite{137}), and proprietary systems (ChatGPT 4o and ChatGPT 4o mini). This diversity ensured comprehensive testing of the framework's applicability across varied domains and complexities.

The validation began with assessing the framework's ability to clarify architectural components by identifying implicit core-agents in existing systems (Section~\ref{sec:evaluation-core-agent-terminology}). For instance, Toolformer and ToolLLM, which do not explicitly define themselves as agents, were analyzed to uncover core-agents, focusing on \textit{Modularity}, \textit{Reusability}, and \textit{Terminological Consistency}. This process confirmed the framework's capacity to delineate components clearly, enhancing \textit{Maintainability} and \textit{Interoperability}. Next, the internal structure of core-agents was evaluated by classifying them as active or passive across the thirteen agents (Section~\ref{sec:evaluation-active-passive-core-agent-internal-structure}). This analysis targeted \textit{Modularity}, \textit{Flexibility}, and \textit{Security}, revealing gaps, such as the absence of security modules in 78\% of tool-using agents (e.g., ToolAlpaca, Gorilla), except in ChemCrow and ChatGPT 4o, which demonstrated robust privacy measures. The classification also highlighted \textit{Scalability}, as passive core-agents (e.g., in Toolformer, Confucius) required less synchronization, unlike active core-agents (e.g., in Chameleon, LLM+P).

The framework's strength in designing scalable and flexible multi-core architectures was tested through five novel agent designs (Section~\ref{sec:evaluation-of-multi-core-agent-architectures}). First, LLM-based Agent 1 (LA1) combined Toolformer and Confucius as a multi-passive core-agent system, assessing \textit{Scalability}, \textit{Reusability}, and \textit{Flexibility}. This design leveraged their complementary tool-handling capabilities, but the framework identified a privacy gap in data transfers, underscoring its ability to pinpoint security needs. Second, LLM-based Agent 2 (LA2) explored ToolLLM and ChatDB integration in two variants: LA2-A (multi-active core-agents) and LA2-B (monolithic active core-agent). This scenario evaluated \textit{Scalability}, \textit{Modularity}, and \textit{Maintainability}, addressing synchronization challenges in LA2-A and potential module conflicts in LA2-B, demonstrating the framework's guidance in resolving architectural complexities. Third, LLM-based Agent 3 (LA3) integrated LLMSafeGuard's security module into ToolLLM, focusing on \textit{Security}, \textit{Privacy}, and \textit{Modifiability}. The seamless incorporation highlighted the framework's support for modular extensions without functional conflicts. Finally, LLM-based Agent 4 (LA4) proposed a hybrid one-active-many-passive architecture, combining LLM+P's active core-agent, LLMSafeGuard's security module, and Toolformer/Confucius passive core-agents. This design assessed \textit{Scalability}, \textit{Flexibility}, \textit{Security}, and robustness, leveraging LLM+P's advanced planning and addressing privacy vulnerabilities in tool usage. By strategically selecting agents with diverse capabilities and designing these scenarios, the LLM-Agent-UMF proved its capacity to guide the creation of modular, secure, and scalable agent architectures, identifying risks and opportunities without requiring implementation.

\section{The LLM-Agent-UMF}
\label{sec:llm-agent-umf}

The LLM-based Agent Unified Modeling Framework provides a modular and scalable architecture for designing LLM-based agents, addressing limitations in existing frameworks by introducing the core-agent as a central coordinator. This section details the framework's components, including the core-agent's internal modules, active/passive classifications, and multi-core architectures, emphasizing \textit{Modularity}, \textit{Security}, and \textit{Flexibility}.

\subsection{Overview}
\label{sec:overview}

The LLM-Agent-UMF redefines LLM-based agent design by introducing the core-agent as the keystone component, orchestrating interactions among Large Language Models (LLMs), tools, and the environment to achieve high-level goals. Inspired by the human brain's modularity, the core-agent comprises five modules---planning, memory, profile, action, and security---designed for \textit{Extensibility} and \textit{Maintainability}, as shown in Figure~\ref{fig:overview-of-core-agent-internal-structure-in-llm-based-agent}. The framework classifies core-agents as active (authoritative, managing complex tasks) or passive (non-authoritative, executing specific actions), enabling versatile architectures like the hybrid one-active-many-passive setup, which balances simplicity and functionality. By adhering to software engineering principles such as the \textit{Single Responsibility Principle} and \textit{Open-Closed Principle}, LLM-Agent-UMF ensures clarity, reusability, and robust security, overcoming the architectural ambiguities and functional overlaps of prior frameworks~\cite{wang2024,cheng2024}.

\subsection{Modeling the Core-Agent Internal Structure}
\label{sec:modeling-core-agent-internal-structure}

Inspired by the human brain's modularity, our framework incorporates five internal modules within the core-agent: planning, memory, profile, action, and security (Figure~\ref{fig:overview-of-core-agent-internal-structure-in-llm-based-agent}). The core-agent acts as the central coordinator, managing interactions and information flow among these modules, as described in the previous section. This modular structure enables independent development and replacement of modules, supporting the addition of new capabilities and adaptation to new technologies. It also facilitates code reusability, allowing modules to be shared across different agents or applications.

\subsubsection{Planning Module}
\label{sec:planning-module}

\begin{figure}[h]
    \centering
    \includegraphics[width=\linewidth]{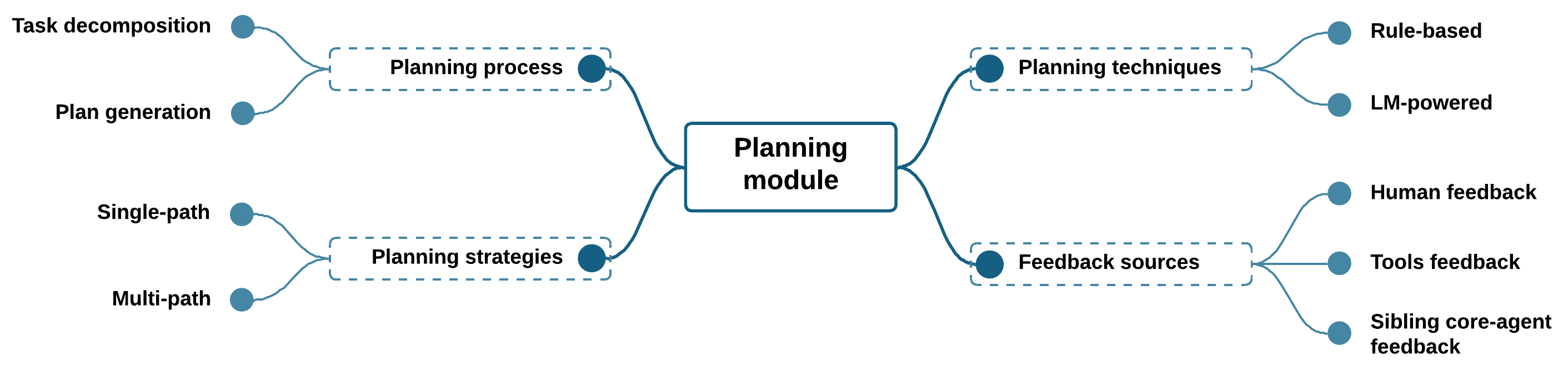}
    % Comment: This figure outlines the functional perspectives of the planning module within the core-agent, covering four aspects: Planning Process (task decomposition and plan generation), Planning Strategies (single-path and multi-path), Planning Techniques (rule-based and LM-powered), and Feedback Sources (human, tool, and sibling core-agent). It emphasizes the module’s role in generating effective plans with LLM support.
    \caption{Planning module functional perspectives}
    \label{fig:planning-module-functional-perspectives}
\end{figure}

The planning module is a pivotal component of the core-agent, enabling the agent to decompose complex problems and generate effective plans. It leverages the LLM's capabilities for understanding nuanced instructions, interpreting implicit information, and adapting to diverse problem domains, producing comprehensive, context-aware plans that enhance the core-agent's decision-making. The module collaborates closely with other core-agent modules, such as the memory module for memory-augmented planning~\cite{108}, utilizing stored information like commonsense knowledge and past experiences. This section details the module's functionalities across four aspects: process, strategies, techniques, and feedback sources (Figure~\ref{fig:planning-module-functional-perspectives}).

\vspace{0.3cm}
\noindent\textbf{Planning process:}

The planning module adopts an incremental approach to generate procedures to undertake, systematically decomposing tasks, unfolding the planning process, and generating and evaluating alternative solutions. This process comprises two main steps~\cite{108}: task decomposition and plan generation.

\begin{itemize}
	\item \textbf{Task decomposition:}

    This initial phase involves breaking down a complex task into simpler subtasks, thereby establishing a structured hierarchy of intermediate goals. The decomposition of complex tasks can adopt two primary approaches: In the non-iterative decomposition approach, the complex task is broken down into simple subtasks all at once. The planning module defines all subtasks, creating a complete task hierarchy in a single step. This method provides a comprehensive overview of the entire task structure upfront. However, the iterative decomposition approach involves a step-by-step breakdown of the task. In fact, the planning module first defines the initial subtask and goal to reach, establish a proper plan and generates the procedures to undertake. After performing the procedures and completing the plan, the output is considered, and the next subtask is defined. This process is repeated, with each new subtask being planned and executed, contributing to defining the next subtask to achieve. The key advantage of this method is its flexibility and ability to adapt the plan based on the outcomes of each subtask. An exemplary application of the iterative approach is the Decomposition-Alignment-Reasoning Agent (DARA) framework {\cite{141}}. DARA demonstrates how iterative decomposition can lead to more precise and context-aware planning, particularly for complex tasks with multiple interdependent subtasks.

    \item \textbf{Plan generation:}

    For each subtask derived from the decomposition step, a specific plan is established outlining the procedures to achieve the task's goal while defining the different tools and parties involved. Depending on the chosen planning strategy, the module can generate either a single plan or multiple candidate-plans for each subtask which will be further detailed in the following section.
\end{itemize}

\vspace{0.3cm}
\noindent\textbf{Planning strategies:}

To guide the planning and execution of complex tasks, the planning module adheres to a specific strategy during the plan generation step. Selecting a strategy influences the effectiveness, efficiency, and robustness of the resulting plans. Within the LLM-Agent-UMF, we define two primary strategies:

\begin{itemize}
    \item \textbf{Single-path strategy:}

    This approach involves generating a singular path or sequence of procedures to achieve the goal, adhering to the plan step-by-step without exploring alternatives, thereby providing a straightforward, deterministic approach to planning. Chain of thought (CoT) {\cite{80}} is an example that outlines such a strategy. Indeed, it uses sequential reasoning as it involves breaking down complex problems into multiple procedures, each built on the previous ones.

	\item \textbf{Multi-path strategy:}

    Consequently, in the single-path strategy, an error in one procedure can render subsequent procedures or the overall plan suboptimal or infeasible~\cite{108}. The multi-path strategy mitigates such failures through two steps: First, the LLM generates multiple plans for the complex task, with each intermediate step offering multiple subsequent paths. Second, the module evaluates and selects the most suitable path.

    As a matter of fact, the Tree of Thoughts (ToT) and Graph of Thoughts (GoT) {\cite{98}} are two frameworks that utilize this multi-path approach. They both operate by leveraging an LLM as a thought generator to produce intermediate procedures, that are structured either as a hierarchical tree in case of ToT or as a more complex graph in case of GoT. However, the complexity of managing these structures cannot be solely handled by the LLM. It requires the integration of a specialized software component responsible for orchestrating the process, interacting with the LLM, and organizing the thoughts into the desired structure, whether a tree or a graph. This essential software component is identified as the planning module within the core-agent. The planning module further evaluates different paths within the generated structure and selects an optimal plan {\cite{108}} based on its assessment.
\end{itemize}

\vspace{0.3cm}
\noindent\textbf{Planning Techniques:}

The planning module follows planning techniques as methodological approaches to construct executable plans. These techniques are chosen based on criteria such as the complexity of the task, and the need for contextual comprehension. Our framework presents two primary techniques:

\begin{itemize}
	\item \textbf{Rule-based technique:}

    Within our architectural framework, rule-based methodologies encompass what is commonly referred to in literature as symbolic planners {\cite{108}}. These techniques proved to be valuable especially in contexts characterized by complex constraints, such as mathematical problem-solving or the generation of plans within highly problematic situations.

    Symbolic planners, leveraging frameworks like PDDL (Planning Domain Definition Language), utilize formal reasoning to delineate optimal trajectories from initial states to targeted goal states {\cite{149}}. These methodologies entail formalizing problem scenarios into structured formats, subsequently subjecting them to specialized planning algorithms.

	\item \textbf{Language model powered technique:}

    Language Model powered (LM-powered) methodologies leverage the vast knowledge and reasoning capabilities inherent in LMs to orchestrate planning strategies. Within our framework, this category also encompasses neural planners~\cite{108}, adept at addressing intricate and vague tasks requiring nuanced comprehension and adaptive problem-solving.
\end{itemize}

This categorization distinguishes LM-driven approaches from rule-based methods, streamlining the understanding of planning techniques within the core-agent architecture.

\vspace{0.3cm}
\noindent\textbf{Feedback Sources:}

Planning without feedback may pose several challenges as feedback plays a crucial role in optimizing the performance of the planning module within the core-agent. For instance, in iterative task decomposition, feedback has an influential impact on the next generated step and enhances the agent's alignment with the user's expectations. To effectively address these challenges, the planning module relies on diverse feedback sources. As outlined in Figure~\ref{fig:overview-of-core-agent-internal-structure-in-llm-based-agent}, the core-agent engages with tools within its system boundaries, as well as entities outside its scope, such as external systems and humans. Consequently, interactions with these components can offer valuable feedback:

\begin{itemize}
	\item \textbf{Human Feedback:}

    Human feedback may be an essential source of information for aligning the planning module with human values and preferences. This feedback results from direct interactions between the core-agent and humans. For example, when the core-agent proposes a plan, humans may provide feedback on its appropriateness, effectiveness, or ethical implications. This feedback could come in various forms, such as ratings, or comments.

	\item \textbf{Tool Feedback:}

    The core-agent often utilizes various tools, which can be internal components of the system or external applications. For instance, an internal calculator tool might raise an exception upon receiving an illegal operation like division by zero. Likewise, external tools provide feedback in the form of error messages, or performance indicators. Indeed, if the core-agent uses a weather prediction remote API, the accuracy of the prediction serves as feedback. This tool-provided feedback helps the core-agent refine its tool selection and usage strategies.

    \item \textbf{Sibling Core-Agent Feedback:}

    In multi-core agent systems, as will be discussed in Section~\ref{sec:multi-active-passive-core-agent-architecture}, feedback from sibling core-agents becomes a valuable source of information. This type of feedback results from interactions and information exchanges between different core-agents within the same system. This intra-agent feedback can include shared observations, alternative perspectives on a problem, or evaluations of proposed plans. Such feedback promotes collaborative problem-solving and allows for cross-validation of plans. It enhances the overall robustness of multi-core agent systems by facilitating collective intelligence.
\end{itemize}

To conclude, the planning module is a critical component of our framework, employing a structured planning process that consists of task decomposition and plan generation steps. This process is distinct from, yet closely intertwined with, the planning strategies---single-path and multi-path---which guide the formulation of plans. These strategies are crucial as they shape the core-agent's approach to problem-solving, influencing both the quality and efficiency of the solutions. By breaking down tasks and generating multiple plans, the core-agent can identify optimal solutions more quickly and effectively.

The planning module's effectiveness is further enhanced by incorporating feedback from various sources, including humans, tools, and sibling core-agents. This feedback loop allows for continuous refinement and adaptation of plans, ensuring that the agent remains responsive and efficient.  Moreover, incorporating planning strategies and feedback mechanisms enhances \textit{Flexibility}, enabling the core-agent to address a wide range of scenarios, from simple tasks to complex challenges. Such strategic diversity equips the system with greater intelligence and versatility.

It is important to note that the planning module does not operate in isolation. It collaborates closely with other modules, particularly the memory module, which will be discussed in the next section. This collaboration, especially in the context of memory-augmented planning, further enhances the core-agent's capabilities by leveraging stored information and past experiences.

\subsubsection{Memory Module}
\label{sec:memory-module}

\begin{figure}[h]
    \centering
    \includegraphics[width=\linewidth]{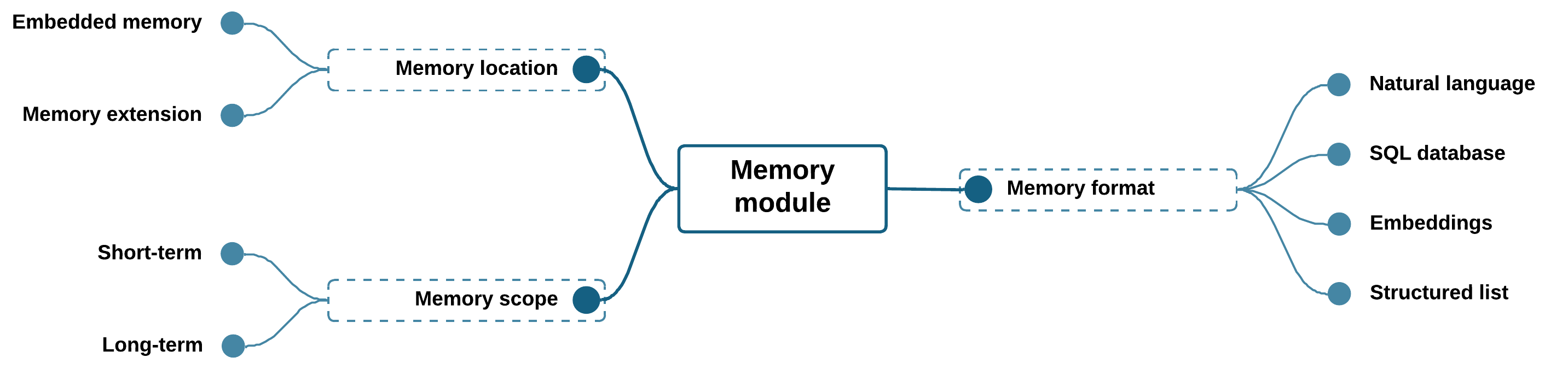}
    % Comment: This figure details the memory module’s functional perspectives, organized into three categories: Memory Scope (short-term and long-term), Memory Location (embedded and extension), and Memory Format (natural language, embeddings, SQL databases, structured lists). It clarifies the module’s role in storing and retrieving information, avoiding overlap with other modules like planning.
    \caption{Memory module functional perspectives}
    \label{fig:memory-module-functional-perspectives}
\end{figure}

The memory module manages the storage and retrieval of information critical to the core-agent's activities, supporting efficient decision-making and task execution. Unlike read-only knowledge repositories, which are handled by the action module (Section~\ref{sec:action-module}), the memory module supports both reading and writing operations to dynamically update task-related data. It is organized around three perspectives: Memory Scope, Memory Location, and Memory Format (Figure~\ref{fig:memory-module-functional-perspectives}).

The \textbf{Memory Scope} perspective categorizes memory into short-term and long-term types, aligned with human memory models~\cite{131}. Short-term memory stores data from one or multiple trials of a specific task~\cite{104}, enabling in-context learning to enhance LLM performance on immediate tasks. Long-term memory retains information over extended periods, maintaining coherence across interactions and adapting from past experiences. For example, MemoryBank~\cite{120} stores user interactions in a symbolic memory, summarizing experiences for future tasks. Due to memory constraints, forgetting mechanisms are employed to selectively retain or discard data~\cite{104}.

The \textbf{Memory Location} perspective distinguishes between Embedded Memory, stored within the core-agent, and Memory Extension, stored externally but within the agent system (Figure~\ref{fig:overview-of-core-agent-internal-structure-in-llm-based-agent}). This categorization supports flexible memory management tailored to architectural needs.

The \textbf{Memory Format} perspective defines how memory is represented, including natural language, embeddings, SQL databases, or structured lists. Natural language offers semantic richness, while embeddings improve retrieval efficiency. For instance, ChatDB~\cite{133} uses SQL databases for structured memory manipulation, and GITM~\cite{127} employs structured lists to organize sequential actions hierarchically. Multiple formats can be combined, as in GITM's key-value lists blending embeddings and natural language, to leverage their respective strengths.

\subsubsection{Profile Module}
\label{sec:profile-module}

\begin{figure}[h]
    \centering
    \includegraphics[width=0.6\linewidth]{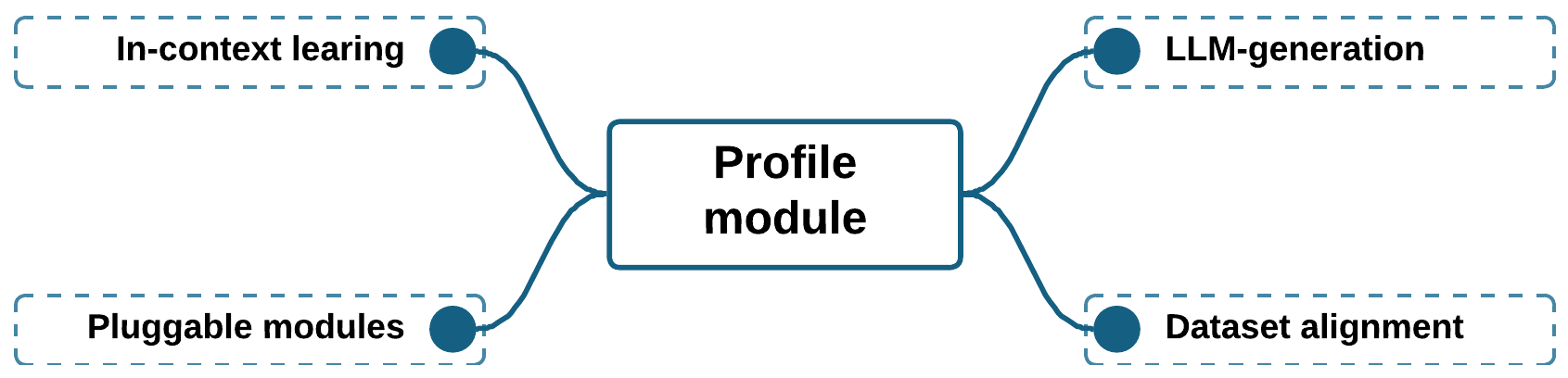}
    % Comment: This figure illustrates the techniques used by the profile module to define the LLM’s role, including Handcrafted In-Context Learning, LLM-Generation, Dataset Alignment, and Fine-tuned Pluggable Modules. It highlights how the module dynamically adapts LLM behavior for specific tasks, enhancing customization and efficiency.
    \caption{Profile module techniques}
    \label{fig:profile-module-techniques}
\end{figure}

The profile module, an integral component of the core-agent's internal structure, defines the role of the Large Language Model (LLM) while clearly distinguishing it from the core-agent's coordinating responsibilities. This separation ensures that the LLM operates with a specific persona tailored to the task, enhancing the core-agent's ability to adapt dynamically to diverse use cases and planning strategies. The module employs four distinct methods to create these profiles, as illustrated in Figure~\ref{fig:profile-module-techniques}: Handcrafted In-Context Learning, LLM-Generation, Dataset Alignment, and Fine-tuned Pluggable Modules.

\textbf{The Handcrafted In-Context Learning Method} enables the core-agent to configure LLM profiles using pre-configured prompts, providing fine-grained control over the LLM's personality and behavior. This approach leverages in-context learning techniques, requiring LLMs that are adept at interpreting such prompts, often necessitating models of substantial size. While straightforward to implement, its reliance on large LLMs can pose computational challenges, making it resource-intensive but effective for precise customization.

\textbf{The LLM-Generation Method} automates profile creation by harnessing the LLM's generative capabilities. It begins by specifying profile characteristics, such as age, gender, or interests, and may incorporate seed profiles as few-shot examples to guide generation. For example, RecAgent~\cite{114} designs prompts that encourage a GPT model to generate comprehensive profile descriptions based on attribute tables, streamlining the process. However, this method's dependence on LLMs introduces risks of biases~\cite{158} or inaccuracies, necessitating careful validation of generated profiles.

\textbf{The Dataset Alignment Method} derives profiles from real-world datasets containing individual data, ensuring realistic and grounded LLM roles. Information from these datasets is organized into natural language prompts that describe the LLM's role characteristics. In a study by~\cite{116}, researchers used GPT-3 with ANES demographic data to assign roles based on attributes like state of residence, mimicking real human behavior reliably. This method's effectiveness relies heavily on the accuracy and representativeness of the datasets, making data quality a critical factor.

\textbf{The Fine-tuned Pluggable Modules Method}, a novel approach, customizes LLM profiles by injecting fine-tuned modules that adjust model behavior~\cite{21}. These modules, trained using Parameter-Efficient Fine-Tuning (PEFT) techniques such as Sequential Adapter, Prompt-tuning, or LoRA~\cite{115}, encode desired profiles directly into model parameters. This eliminates the need for extensive in-context learning prompts, reducing context size and memory footprint while maintaining flexibility and performance, offering an efficient and precise customization solution.

\subsubsection{Action Module}
\label{sec:action-module}

\begin{figure}[h]
    \centering
    \includegraphics[width=\linewidth]{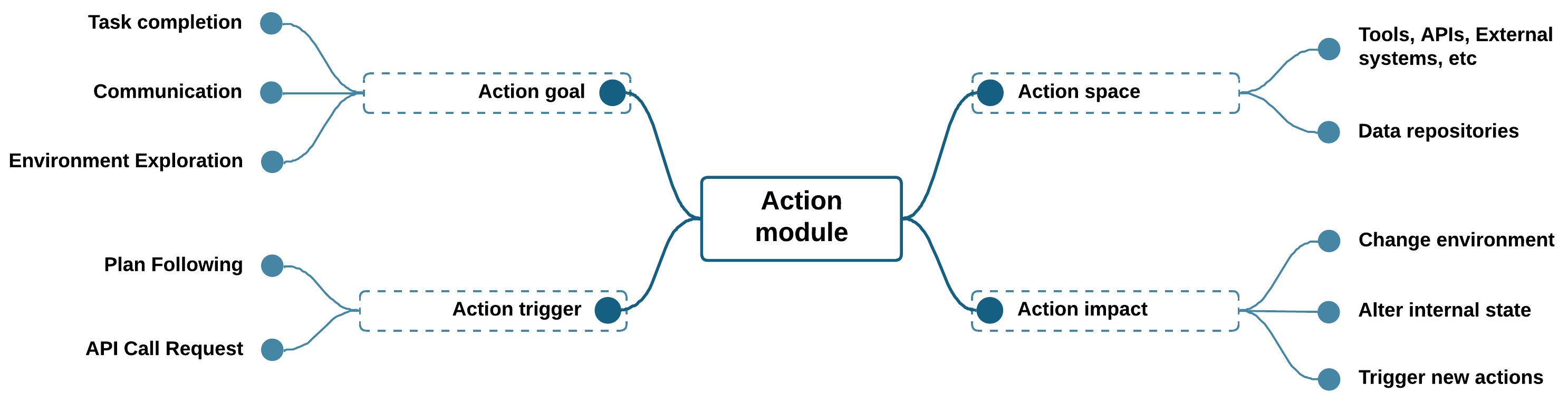}
    % Comment: This figure depicts the functional perspectives of the action module, covering Action Goal (task completion, communication, environment exploration), Action Trigger (plan following, API call request), Action Space (possible actions via tools or data sources), and Action Impact (environmental or internal state changes). It emphasizes the module’s role in executing high-level plans securely.
    \caption{Action module functional perspectives}
    \label{fig:action-module-functional-perspectives}
\end{figure}

The action module, a core component of the core-agent, translates high-level instructions from the planning module into executable low-level actions by leveraging tools within its environment. It collaborates closely with the security module to ensure that actions adhere to predefined integrity, safety and privacy criteria, as discussed in Section~\ref{sec:security-module}. The module's operations are structured around four key perspectives: Action Goal, Action Trigger, Action Space, and Action Impact, as depicted in Figure~\ref{fig:action-module-functional-perspectives}.

The \textbf{Action Goal} perspective outlines the objectives the core-agent aims to achieve through its actions. These objectives include Task Completion, such as solving a specific problem; Communication, such as interacting with users or other systems; and Environment Exploration, such as navigating a virtual or physical space. This perspective ensures that each action is purposefully aligned with the core-agent's intended outcomes, providing a clear direction for the execution process.

The \textbf{Action Trigger} perspective identifies the catalysts that initiate actions, encompassing two primary triggers: Plan Following and API Call Request. \textbf{Plan Following} involves executing actions based on pre-generated plans from the planning module. For instance, in GITM~\cite{127}, the agent decomposes tasks into subtasks and executes actions in the Minecraft world via an LLM interface. The \textbf{API Call Request} trigger, inspired by techniques like TALM~\cite{126} and ToolFormer~\cite{22}, enables actions in response to LLM-initiated API calls, facilitating seamless integration with external resources.

The \textbf{Action Space} perspective defines the set of possible actions that can be performed by the core-agent in response to its objectives and environmental factors. This concept is distinct from operations involving internal state changes and memory management, which are handled separately by the Memory Module. The action module can leverage tools ranging from basic software components like calculators to more advanced systems accessed through API calls, as exemplified in the HuggingGPT work that leverages the HuggingFace API to accomplish complex user tasks~\cite{7}. Alternatively, the core-agent may expand its operational scope and knowledge by communicating with read-only data sources such as knowledge repositories. This process is often referred to as Retrieval-Augmented Generation (RAG)~\cite{165}. This is particularly crucial for AI systems deployed in critical industries like healthcare where explainability in decision-making processes becomes paramount~\cite{164, 167}. By interacting with these external resources, transparency, provenance, and traceability are ensured to guarantee the reliability of the agent's outputs~\cite{166}. It is useful to note here that data repositories can have the same formats discussed in the memory module (Section~\ref{sec:memory-module}). For instance, LlamaIndex~\cite{162} stores data as vector embeddings at the indexing stage to leverage semantic search, where the similarity between embeddings is used to rank documents by their relevance to a query. In contrast, ReAct~\cite{123} uses a textual data repository, like Wikipedia, to mitigate error propagation in chain-of-thought reasoning.

The \textbf{Action Impact} perspective captures the consequences of actions, which may include altering the environment, such as moving locations, gathering resources, or synthesizing chemical compounds~\cite{132}. Actions can also modify the core-agent's internal state, particularly when acquiring new knowledge. This perspective highlights the tangible and intangible effects of actions, underscoring their role in achieving the core-agent's goals and adapting to dynamic contexts.

\subsubsection{Security Module}
\label{sec:security-module}

\begin{figure}[h]
    \centering
    \includegraphics[width=\linewidth]{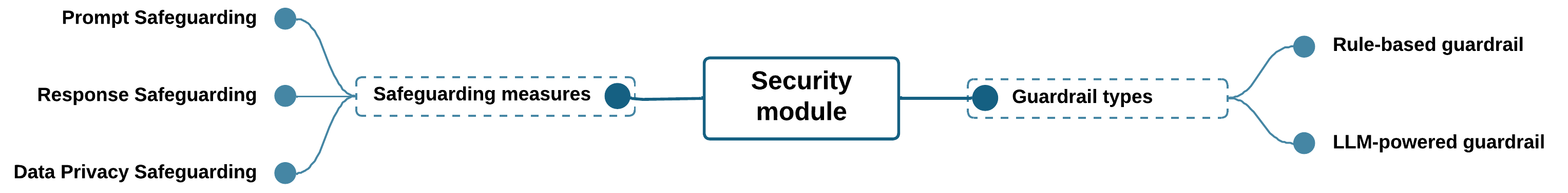}
    \caption{Security module functional perspectives}
    \label{fig:security-module-functional-perspectives}
\end{figure}

The security module addresses risks in Large Language Model (LLM) deployment, such as unauthorized or unethical use, data biases, privacy breaches, and misinformation~\cite{168, 135}. It defines \textit{security} as specifying \textit{what is permitted}, \textit{by whom}, \textit{when}, and \textit{how}, serving as the core-agent's enforcement mechanism, while \textit{safety} focuses on preventing harm. Aligned with standards like the \textit{EU AI Act}~\cite{eu_ai_act} and \textit{ISO/IEC 22989:2022}~\cite{iso_22989_2022}, which emphasize robustness and cybersecurity, the module continuously monitors the action module to enforce security and safety requirements. Incorporating \textit{guardrails}---specialized algorithms that detect and mitigate misuse~\cite{111}---it ensures robust protection for LLM-based systems. The module integrates the Confidentiality, Integrity, and Availability (C.I.A.) triad~\cite{151}, operationalizing these principles through two key perspectives: Safeguarding Measures (Prompt, Response, and Data Privacy Safeguarding) and Guardrail Types (rule-based and LLM-based guardrails), as illustrated in Figure~\ref{fig:security-module-functional-perspectives}, operating independently of cognitive functions like planning and memory to maintain robust protection.

\vspace{0.3cm}
\noindent\textbf{Security Measures:}

The security module implements three safeguarding mechanisms---Prompt Safeguarding, Response Safeguarding, and Data Privacy Safeguarding. These mechanisms, applied at input and output levels, form a layered defense to mitigate threats like prompt injections and data breaches.

\begin{itemize}
    \item \textbf{Prompt Safeguarding:}

    Prompt safeguarding necessitates employing measures to detect and mitigate unauthorized access to Large Language Models (LLMs) through prompt injection attacks {\cite{142}}. Techniques for enhancing the security of LLM-based agents can be integrated directly into the LLM itself, with Adversarial Training (AT) being a prominent example {\cite{129}}. AT enhances an LLM's defense mechanisms by fine-tuning it with augmented training data containing adversarial examples, thereby increasing the model's ability to safeguard against malicious prompts and improving its robustness. However, AT faces significant limitations, such as challenges in efficiently selecting adversarial examples and vulnerability to adversarial perturbations like HOUYI {\cite{143}}, a black-box prompt injection attack. Moreover, such training-based security techniques may impact the generative performance of the LLM, necessitating additional evaluation steps.

    To address these limitations and enhance protection, other techniques decouple security measures from the LLM, delegating them to a distinct entity identified as a core-agent supplemented with a dedicated security module, as illustrated in Figure~\ref{fig:overview-of-core-agent-internal-structure-in-llm-based-agent}. This decoupling is critical for robust prompt safeguarding and enables the implementation of advanced security protocols that continuously monitor and filter LLM inputs, evolving independently of the LLM's training process. An example of this approach is Nvidia NeMo {\cite{144}}, which functions as an intermediary layer between users and LLMs, employing advanced techniques such as vector databases and comparison with stored canonical forms to detect and mitigate malicious inputs before they reach the model, thereby providing robust prompt safeguarding without directly modifying the underlying LLM.

    \item \textbf{Response Safeguarding:}

    The recent survey~{\cite{112}} has demonstrated that despite the implementation of prompt safeguarding techniques, the overall resilience of Large Language Models (LLMs) against advanced attacks, known as jailbreaks, may not experience significant improvement. These jailbreaks are designed to exploit biases or vulnerabilities within language models by manipulating their responses. Notable examples include white-box attacks AutoDAN-Zhu {\cite{145}}, which generate stealthy prompts to avoid triggering the model protective mechanisms. Additionally, black-box attacks leverage manually crafted prompts to deceive the LLM {\cite{112}}. The effectiveness of these jailbreak techniques is further illustrated in~{\cite{111}}, where researchers successfully achieved a jailbreak attack on ChatGPT 3 by framing potentially harmful query, ``how to hotwire a car'' as a hypothetical scenario. The existence of these jailbreak methods highlights the urgent necessity for continuous and rigorous monitoring of LLM outputs to detect and mitigate potential breaches. It is useful to note that the aforementioned jailbreak attack was addressed and resolved in later versions of ChatGPT, 3.5 and 4, as confirmed by~{\cite{111}}.

    Guarding the agent outputs is considered a supplementary measure alongside prompt safeguarding, with the objective of guaranteeing the safety, integrity and authenticity of the text generated by LLMs. This includes detecting and redacting harmful content while maintaining coherence and relevance. Two notable examples of such techniques are Guardrails AI {\cite{112}} and LLMSafeGuard {\cite{137}}, discussed further in Section~\ref{sec:evaluation}.

    \item \textbf{Data Privacy Safeguarding:}

    Lastly, safeguarding data privacy is pivotal, especially when handling sensitive or personal information. It is essential to ensure that LLMs are protected against sensitive data breaches {\cite{112}}. Existing research has predominantly focused on securing training data through traditional techniques such as Differential Privacy {\cite{146}} and watermarking {\cite{147}}. As a matter of fact, Differential Privacy tuned models add noise to the data, making it difficult to identify individual data points. Similarly, watermarking techniques embed identifiable markers into LLM outputs, allowing for the tracing of data origin and preventing unauthorized use.

    However, under our proposed framework, the primary objective shifts from protecting the privacy of training data to ensuring that the LLM does not divulge sensitive information to external tools. This approach aims to maintain the privacy and security of data while interacting with other systems. Indeed, as previously mentioned, the core-agent can leverage external tools, APIs, and knowledge repositories to augment the capabilities of the LLM. Being part of the agent as a whole system, internal tools are part of the privacy circle, thereby, there is no need to apply security measures on communication procedures. However, the use of external resources introduces potential risks that require specific attention. These risks include a lack of robust data privacy measures, potentially leading to data leaks or unauthorized access. For example, when the core-agent utilizes third-party services to access additional information or perform specific tasks, the data transmitted may contain sensitive details that specific systems are not authorized to access. Additionally, such sensitive information could be intercepted or mishandled if proper secure channels are not leveraged.

    To mitigate these risks, the security module within the core-agent must implement a range of powerful techniques such as access control mechanisms, and data encryption. Therefore, the framework ensures that interactions with external resources maintain the highest standards of security and data privacy.
\end{itemize}

The security module operates across three axes: Prompt Safeguarding, Response Safeguarding, and Data Privacy Safeguarding. These integrated mechanisms mitigate threats like prompt injection and data breaches, providing layered protection at input and output levels. This approach prioritizes security in the core-agent's operations, from data retrieval to external communication.

\vspace{0.3cm}
\noindent\textbf{Guardrail Types:}

To implement the security axes, the security module uses guardrail methodologies as its operational layer, translating security objectives into actionable safeguards. Two primary types are rule-based guardrails and LLM-based guardrails, as outlined in {\cite{112}}.

\begin{itemize}
    \item \textbf{Rule-based guardrails:}

    These guardrails operate based on a predetermined set of rules and regulations aimed at screening and preventing potentially detrimental or undesirable inputs/outputs from LLMs. To elucidate the process, users define the content necessitating protection. Subsequently, the guardrails assess the inputs/outputs against these predefined regulations, and custom rules {\cite{112}}, to ascertain compliance. In instances where the content is deemed unsafe, it may be obstructed, or a cautionary alert may be issued. For instance, the Adversarial Robustness Toolbox (ART) {\cite{152}} is specifically designed to bolster the security and robustness of models against adversarial attacks. It offers tools and methods to defend against and adapt to malicious inputs, thereby safeguarding AI applications from potential vulnerabilities.

    \item \textbf{LLM-powered guardrails:}

    While rule-based guardrails provide a solid foundation for safeguarding LLM operations, they face limitations in adaptability and maintenance. The need for manual, continuous improvement and intervention to upgrade rules can be time-consuming and may struggle to keep pace with rapidly evolving threats and diverse use cases. LLM-powered guardrails offer a compelling solution to these challenges. A prevalent design approach for constructing these guardrails involves the usage of neural-symbolic agents {\cite{112}}. These agents, functioning akin to core-agents from a security standpoint, undertake the critical task of analyzing input and output, ensuring their adherence to a predefined set of requirements. By leveraging the inherent learning and adaptability capabilities of language models, these guardrails can evolve and respond to new situations in a faster and more automated manner. They can understand context, nuance, and intent more effectively than rigid rule sets, allowing more sophisticated and flexible protection mechanisms.

    Moreover, neural-symbolic agents resolve conflicts that may arise between requirements, leverage historical data to reason symbolically and possess the capability to collaborate with other AI systems {\cite{112}}. While LLM-based solutions may introduce computational overhead, this potential drawback can be mitigated by adopting lightweight models specifically designed for guardrail tasks. These optimized models can provide the benefits of LLM-powered security with reduced resource demands, striking a balance between robust protection and operational efficiency.

    In fact, within the general scope of the agent, the core-agent can communicate with an auxiliary LLM, which is finetuned on specialized dataset to set the acceptability guidelines of the response generated by the main LLM. The core-agent allows for customization of guardrail rules, including monitoring and enforcement protocols {\cite{111}}. These customized rules are then passed to the auxiliary LLM to classify the nature of the input. Such classification helps the core-agent to decide whether the requirements are fulfilled or not.  A leading example of this approach is LLaMA Guard {\cite{111}}. Introduced by Meta (Facebook), it was designed specifically to guarantee the security and reasonable utilization of LLaMA models and used to analyze both input and output data. It employs predictive classification techniques to assess and improve security across user-specified categories. This implementation underscores the critical role of LLM-based guardrails in fortifying the integrity and reliability of AI systems. By leveraging the LLM's capabilities to understand and enforce complex security rules, Llama Guard provides a flexible and powerful mechanism for ensuring safe and responsible AI operation, particularly in next-generation LLM models like Llama 3.1 {\cite{161}}.
\end{itemize}

\subsection{Active/Passive Core-Agent Classification}
\label{sec:active-passive-core-agent-classification}

As discussed in the preceding sections, the core-agent represents a distinct entity within the LLM-Agent-UMF. While the LLM excels in cognitive tasks such as understanding, reasoning, and generating responses, it lacks the capability to directly interact with the environment or external tools. This is where the core-agent plays a crucial role. It bridges the gap between the LLM's cognitive abilities and the need to engage with external sources, enabling seamless integration with various tools and systems. The core-agent is thus characterized by its action capabilities and its ability to respond to user requests through interaction with these diverse tools.
In fact, ToolLLM {\cite{1}} is a general tool-use framework that enhances LLMs capabilities enabling agents to use external tools and APIs. It uses a neural API retriever to recommend appropriate APIs for each instruction. Then they employ a depth-first search-based decision tree algorithm to evaluate multiple reasoning traces and expand the search space. Consequently, it enhances the planning ability of the retriever and empowers the finetuned LLM, ToolLlaMA, to generate adequate instructions. The retriever here, in association with the search-based decision tree algorithm, satisfies our definition of a core-agent. In this case where the LLM-based agent conducts cognitive tasks, memory and planning modules are essential in the core-agent to ensure reasoning capabilities because they enable the agent to retain and recall past experiences, plan and synchronize actions, reason and make decisions {\cite{104,108}}.

In other cases, such as Toolformer {\cite{22}}, we identify entities that fit our definition of a core-agent but lack both planning and memory modules. Indeed, Toolformer fine-tunes its LLM on function calling, enabling it to generate API requests within natural language as needed. Consequently, the LLM determines when to make an API call, which API to use, and how to integrate the results, while the actual execution of the API request is delegated to an entity that we identify as a core-agent. In this case, the planning module of the core-agent is obsolete because planning is handled solely by the LLM. Yet, its action module is present because it is still responsible for executing API calls systematically. For example, if the model suggests using a calculator API, the core-agent retrieves the arguments for the mathematical operation, performs the calculation, and returns the computed result to the LLM.

The inspection of the state-of-the-art led to the conclusion that the action module is always indispensable in a core-agent as it is responsible for producing the executive steps to achieve their goals. However, the architectural disparities in core-agents and the absence of some modules in some proposed agent systems highlight the need to introduce a new taxonomy classifying core-agents into two distinct categories: Active core-agents (Figure~\ref{fig:overview-of-core-agent-internal-structure-in-llm-based-agent}) and passive core-agents (Figure~\ref{fig:llm-agent-with-passive-core-agent}). The following sections analyze and pinpoint the main differences and similarities between active and passive core-agents in their structural alignment with our framework.

\subsubsection{Active Core-Agents}
\label{sec:active-core-agent}

Active core-agents encompass all five modules described in Section~\ref{sec:modeling-core-agent-internal-structure} and illustrated in Figure~\ref{fig:overview-of-core-agent-internal-structure-in-llm-based-agent}, but what differentiates an active from a passive core-agent is its managerial aspect. An active core-agent is characterized by its leading position in the agent as the orchestrator of other components, so naturally, it requires a planning module to divide tasks into subtasks and collaborates with the memory module to provide the necessary context, analyze information, and make decisions. Consequently, we consider an active core-agent to be stateful, meaning it can maintain information about its past interactions and states over time. This is facilitated by an adaptive memory that captures and stores various aspects of the agent's lifecycle, allowing it to use this historical data to inform future actions and decisions. The profile module role is emphasized in the active core-agent category, because it guides the LLM's behavior in a certain direction. Furthermore, the security module plays a prominent role in safeguarding the communication between the LLM and the human, ensuring a reliable exchange; Acting as an intermediary, the core-agent safeguards the LLM from threats such as jailbreak attempts and protects user data privacy by implementing safety measures as outlined in Table~\ref{tab:table1}.

As highlighted in~\cite{cheng2024}, active core-agents are more effective because they incorporate planning and memory modules, which enable them to reason, plan, and execute tasks efficiently. This structure allows the agent to adapt to changing situations and make informed decisions, making the system more robust and capable.

However, relying solely on active core-agents would increase the complexity of the agent, which can lead to \textit{Scalability} issues and negatively impacts the \textit{Maintainability} of the agent as it will hinder and complicate future improvement efforts. As noted by~\cite{cheng2024}, ``the complexity of the agent system grows exponentially with the number of tasks it needs to perform''. Therefore, rather than centralizing responsibilities on one entity, it would be more beneficial and adhering to the \textit{Single Responsibility Principle}, if we leverage other core-agents to granulate task execution and reduce the complexity of the agent system. This approach is supported by the concept of \textit{Separation of Concerns} in software engineering, which emphasizes the importance of dividing responsibilities among multiple components to improve system \textit{Modularity} and \textit{Maintainability}. By distributing tasks among multiple core-agents, we can reduce the cognitive load on individual core-agents, improve system efficiency, and enhance overall performance.

\subsubsection{Passive Core-Agents}
\label{sec:passive-core-agent}

\begin{figure}[h]
    \centering
    \includegraphics[width=\linewidth]{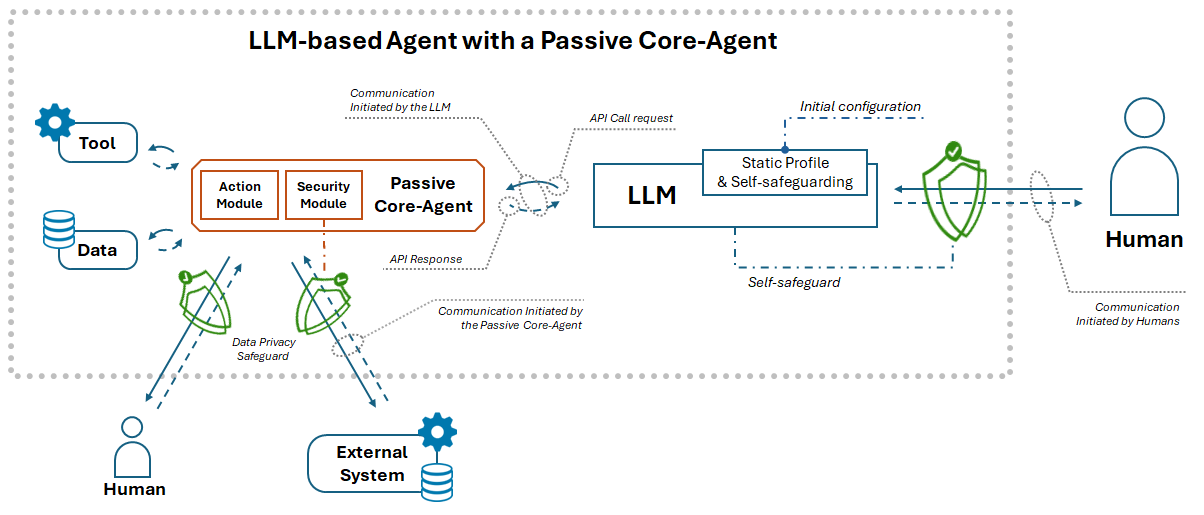}
    % Comment: This figure illustrates an LLM-based agent architecture featuring a passive core-agent, which lacks authoritative control and primarily executes LLM-directed actions via the action module. It shows minimal module presence, emphasizing simplicity and reliance on the LLM for decision-making. It also emphasize the present of a security module handling Data Privacy Safeguarding.
    \caption{LLM-based agent architecture including a passive core-agent}
    \label{fig:llm-agent-with-passive-core-agent}
\end{figure}

Passive core-agents are employed when LLMs cover all cognitive tasks of the agent such as planning and taking decisions, while passive core-agent's role is mainly to execute specific procedures. As a direct consequence, the planning module becomes unnecessary and likewise the memory needed in reasoning. Unlike active core-agents, passive core-agents are stateless, and the short-term memory is handled by the LLM, covering only the current task's state. In LLM-based agents, passive core-agents, which always follow instructions from domain-specific LLMs, lack the ability to control the profile of the LLM thus do not possess a Profile module. The LLM profile may be statically defined during the system setup or dynamically defined by another entity, which will be discussed in the next section.

\begin{table}[h]
    \caption{Active and passive core-agents internal structure}
    \label{tab:table1}

    \centering
    \hspace*{-0.5cm}
    \begin{tabular}{|c|ll|c|c|}

    \hline
    \multicolumn{3}{|c|}{\textbf{Core-Agent Structure}} & & \\
    \cline{1-3}
    \textbf{Modules} &
      \multicolumn{2}{l|}{\textbf{Sub-modules / Methods}} &
      \multirow{-2}{*}{\textbf{\begin{tabular}[c]{@{}c@{}}Active\\Core-Agent\end{tabular}}} &
      \multirow{-2}{*}{\textbf{\begin{tabular}[c]{@{}c@{}}Passive\\Core-Agent\end{tabular}}} \\ \hline

    % Planning module
    \multicolumn{1}{|l|}{} &
      \multicolumn{1}{l|}{} &
      Rule-based & & \\
      \cline{3-3}
    \multicolumn{1}{|l|}{} &
      \multicolumn{1}{l|}{\multirow{-2}{*}{Planning techniques}} &
      LM-powered & & \\
      \cline{2-3}
    \multicolumn{1}{|l|}{} &
      \multicolumn{1}{l|}{} & Task decomposition & & \\
      \cline{3-3}
    \multicolumn{1}{|l|}{} &
      \multicolumn{1}{l|}{\multirow{-2}{*}{Planning process}} &
      Plan generation & & \\
      \cline{2-3}
    \multicolumn{1}{|l|}{} &
      \multicolumn{1}{l|}{} & Single-path strategy & & \\
      \cline{3-3}
    \multicolumn{1}{|l|}{} &
      \multicolumn{1}{l|}{\multirow{-2}{*}{Planning strategies}} &
      Multi-path strategy & & \\
      \cline{2-3}
    \multicolumn{1}{|l|}{} &
      \multicolumn{1}{l|}{} & Human feedback & & \\
      \cline{3-3}
    \multicolumn{1}{|l|}{} &
      \multicolumn{1}{l|}{} & Tools feedback & & \\
      \cline{3-3}
    \multicolumn{1}{|c|}{\multirow{-10}{*}{\textbf{Planning}}} &
      \multicolumn{1}{l|}{\multirow{-3}{*}{Feedback sources}} &
      Sibling core-agent feedback & \multirow{-10}{*}{X} & \multirow{-10}{*}{} \\
      \hline

    % Memory module
    \multicolumn{1}{|l|}{} &
      \multicolumn{1}{l|}{} & Embedded memory & & \\
      \cline{3-3}
    \multicolumn{1}{|l|}{} &
      \multicolumn{1}{l|}{\multirow{-2}{*}{Memory location}} &
      Memory extension & & \\
      \cline{2-3}
    \multicolumn{1}{|l|}{} &
      \multicolumn{1}{l|}{} & Short-term & & \\
      \cline{3-3}
    \multicolumn{1}{|l|}{} &
      \multicolumn{1}{l|}{\multirow{-2}{*}{Memory scope}} & Long-term & & \\
      \cline{2-3}
    \multicolumn{1}{|l|}{} &
      \multicolumn{1}{l|}{} & Natural language & & \\
      \cline{3-3}
    \multicolumn{1}{|l|}{} &
      \multicolumn{1}{l|}{} & SQL database & & \\
      \cline{3-3}
    \multicolumn{1}{|l|}{} &
      \multicolumn{1}{l|}{} & Embeddings & & \\
      \cline{3-3}
    \multicolumn{1}{|c|}{\multirow{-8}{*}{\textbf{Memory}}} &
      \multicolumn{1}{l|}{\multirow{-4}{*}{Memory format}} &
      Structured list & \multirow{-8}{*}{X} & \multirow{-8}{*}{} \\
      \hline

    % Profile module
    \multicolumn{1}{|l|}{} &
      \multicolumn{2}{l|}{Handcrafted in-context learning method} & X & {[}*{]} \\
      \cline{2-5} 
    \multicolumn{1}{|l|}{} &
      \multicolumn{2}{l|}{Fine-tuned pluggable modules method} & X & {[}*{]} \\
      \cline{2-5} 
    \multicolumn{1}{|l|}{} &
      \multicolumn{2}{l|}{LLM-generation method} & X & \\
      \cline{2-5} 
    \multicolumn{1}{|c|}{\multirow{-4}{*}{\textbf{Profile}}} &
      \multicolumn{2}{l|}{Dataset alignment method} & X & \\
      \hline

    % Action module
    \multicolumn{1}{|l|}{} &
      \multicolumn{1}{l|}{} & Task completion & X & X \\
      \cline{3-5} 
    \multicolumn{1}{|l|}{} &
      \multicolumn{1}{l|}{} &
      Communication & X {[}**{]} & X {[}***{]} \\
      \cline{3-5} 
    \multicolumn{1}{|l|}{} &
      \multicolumn{1}{l|}{\multirow{-3}{*}{Action goals}} &
      Environment Exploration & X & \\
      \cline{2-5}
    \multicolumn{1}{|l|}{} &
      \multicolumn{1}{l|}{} &
      Plan Following & X & \\
      \cline{3-5}
    \multicolumn{1}{|l|}{} &
      \multicolumn{1}{l|}{\multirow{-2}{*}{Action trigger}} &
      API Call Request & X & X \\
      \cline{2-5}
    \multicolumn{1}{|l|}{} &
      \multicolumn{1}{l|}{} &
      Tools (APIs, External systems, etc) & X & X \\
      \cline{3-5} 
    \multicolumn{1}{|l|}{} &
      \multicolumn{1}{l|}{\multirow{-2}{*}{Action space}} &
      Data repositories & X & X \\
      \cline{2-5}
    \multicolumn{1}{|l|}{} &
      \multicolumn{1}{l|}{} & Change environment & X & X \\
      \cline{3-5}
    \multicolumn{1}{|l|}{} &
      \multicolumn{1}{l|}{} & Alter internal state & X & \\
      \cline{3-5}
    \multicolumn{1}{|c|}{\multirow{-11}{*}{\textbf{Action}}} &
      \multicolumn{1}{l|}{\multirow{-3}{*}{Action impact}} & Trigger new actions & X & X \\
      \hline

    % Security module
    \multicolumn{1}{|l|}{} &
      \multicolumn{1}{l|}{} &
      Prompt Safeguarding & X & \\
      \cline{3-5}
    \multicolumn{1}{|l|}{} &
      \multicolumn{1}{l|}{} &
      Response Safeguarding & X &
       \\ \cline{3-5}
    \multicolumn{1}{|l|}{} &
      \multicolumn{1}{l|}{\multirow{-3}{*}{Safeguarding measures}} &
      Data Privacy Safeguarding & X & X \\
      \cline{2-5}
    \multicolumn{1}{|l|}{} &
      \multicolumn{1}{l|}{} &
      Rule-based guardrail & X & X \\
      \cline{3-5}
    \multicolumn{1}{|c|}{\multirow{-5}{*}{\textbf{Security}}} &
      \multicolumn{1}{l|}{\multirow{-2}{*}{Guardrail types}} &
      LLM-powered guardrail & X & X \\
      \hline
    \end{tabular}

    % Footer
    \begin{flushleft}
    \footnotesize{$^{[*]}$ Passive core-agents do not have a profile module. Depending on the architecture of the whole agent, the LLM' s profile will be set either statically or dynamically, but not by the passive core-agent as it has no control over it.} \\
    \footnotesize{$^{[**]}$ Communication can be initiated by either Humans or the active core-agent.} \\
    \footnotesize{$^{[***]}$ Communication can only be initiated by the passive core-agent.}
    \end{flushleft}

\end{table}

The most essential module in a passive core-agent is the action module. Our framework posits that the function of a passive core-agent is limited to specific task execution. Actions are often triggered by API call requests, which are not decision-based nor self-generated by the passive core-agent but provided by another entity (e.g., LLM or an active core-agent) as shown in Figure~\ref{fig:llm-agent-with-passive-core-agent}. The actions do not alter the internal state of the agent or change a predetermined plan. This again points to the absence of a planning module in passive core-agents. Furthermore, we introduce another distinction between passive and active core-agents: the communication between humans and core-agents is interactive and bidirectional in both categories, aiming to gather information and/or feedback. However, as pointed out by Figure~\ref{fig:llm-agent-with-passive-core-agent}, the communication between passive core-agents and humans can only be initiated from the passive core-agent part, unlike active core-agents, where communication can be initiated by either party.

Despite not being directly responsible for handling prompts from humans and providing generated text responses, passive core-agents should still possess a robust security module. This component is crucial in ensuring privacy during their interactions with other humans or third-party systems by preventing leakage of sensitive data while minimizing potential threats and breaches. Consequently, this bolsters the overall trustworthiness and reliability of LLM-based agent applications. It is also important to note that in this setup, as illustrated in Figure~\ref{fig:llm-agent-with-passive-core-agent}, the LLM ensures by itself the safety of the prompts by implementing one of the mechanisms previously discussed in Section~\ref{sec:security-module} such as adversarial training.

Describing the characteristics of both active and passive core-agent classes results in constructing a well-structured summary, Table~\ref{tab:table1}, that encapsulates the distinct modules, their respective sub-modules, and the underlying methods for each category. This enables a comprehensive understanding of their functional differences and similarities within the context of LLM-based agents.

\begin{table}[h]
    \caption{Advantages and disadvantages of active and passive core-agents}
    \label{tab:table2}

    \centering
    % \hspace*{-0.5cm}
    \begin{tabular}{|lp{9.5cm}|c|c|}
    \hline
    \multicolumn{2}{|l|}{
      \textbf{Core-agents' advantages and disadvantages}} &
      \multicolumn{1}{c|}{\textbf{\begin{tabular}[c]{@{}c@{}}Active\\Core-Agent\end{tabular}}} &
      \multicolumn{1}{c|}{\textbf{\begin{tabular}[c]{@{}c@{}}Passive\\Core-Agent\end{tabular}}} \\
      \hline
    \multicolumn{1}{|l|}{\multirow{12}{*}{\textbf{Advantages}}} &
    Reinforce Single Responsibility Principle (SRP). & & X \\
    \cline{2-4}
    \multicolumn{1}{|l|}{} &
    Imply simple implementation based on two modules.  & & X \\
    \cline{2-4}
    \multicolumn{1}{|l|}{} & Improve modularity and reduce complexity of   the system.   & X & X \\
    \cline{2-4}
    \multicolumn{1}{|l|}{} & Improve component reusability. & X & X \\
    \cline{2-4}
    \multicolumn{1}{|l|}{} &
    Improve composability and integration in   multi core-agents' setups without any (or with minor) synchronization. & & X \\
    \cline{2-4}
    \multicolumn{1}{|l|}{} & Enhance LLM planning and memory capabilities. & X & \\
    \cline{2-4}
    \multicolumn{1}{|l|}{} & Handle complex tasks. & X & \\
    \cline{2-4}
    \multicolumn{1}{|l|}{} & Access to memory and contextual data. & X & \\
    \cline{2-4}
    \multicolumn{1}{|l|}{} & Possess flexible profile that can be adapted dynamically. & X & \\ \cline{2-4}
    \multicolumn{1}{|l|}{} & Break down complex tasks into subtasks. & X & \\
    \cline{2-4}
    \multicolumn{1}{|l|}{} & Possess multi-tasking capabilities. & X & X \\
    \cline{2-4}
    \multicolumn{1}{|l|}{} & Protects against adversarial attacks. & X & \\
    \hline
    \multicolumn{1}{|l|}{} & Imply complex implementation with multiple modules. & X & \\
    \cline{2-4}
    \multicolumn{1}{|l|}{\multirow{3}{*}{\textbf{Disadvantages}}}
    & Synchronization is needed in multi core-agents' setups. & X & \\
    \cline{2-4}
    \multicolumn{1}{|l|}{} & Handle tasks with limited complexity. & & X \\
    \cline{2-4}
    \multicolumn{1}{|l|}{} & Preclude human-initiated communication& & X \\
    \cline{2-4}
    \multicolumn{1}{|l|}{} & Lacks memory, limiting visibility into the agent's overall status. & & X \\
    \hline
    \end{tabular}

\end{table}

In this section, we detailed the construction of passive and active core-agents, emphasizing their architectural design. This analysis allowed us to identify their utilities and limitations, which are detailed in Table~\ref{tab:table2}. Both categories improve modularity, reduce system complexity, and possess multi-tasking capabilities. Passive core-agents reinforce the \textit{Single Responsibility Principle} and imply simple implementation based only on action and security modules, enhancing reusability and offering straightforward integration into multi core-agents' setups with minimal synchronization requirements which will be discussed in the Section~\ref{sec:multi-active-passive-core-agent-architecture}. However, their simplicity limits their ability to handle complex tasks, precludes human-initiated communication, and lacks memory. Thus, it restricts visibility into the agent's overall status and contextual data access, which is only available via API call requests. Additionally, they have no control over the LLM profile.

In contrast, active core-agents enhance LLM planning and memory capabilities, making them suitable for handling complex tasks. They can access memory and contextual data, control dynamically LLM's profile, and break down complex tasks into manageable subtasks. Despite these advantages, active core-agents require complex implementation involving extra modules compared to passive core-agent and intricate synchronization in multi core-agent' setups, which will be detailed in Section~\ref{sec:multi-active-passive-core-agent-architecture}.

\subsection{Multi Active/Passive Core-Agent Architecture}
\label{sec:multi-active-passive-core-agent-architecture}

Handling complex tasks often necessitates the use of multiple agents, as a single agent may not possess the requisite capabilities or expertise to tackle diverse domains. However, LLM-based multi-agent systems face considerable challenges, including scalability, integration, management of inter-agent relationships, and ensuring interpretability in managing intricate tasks {\cite{40}}.
In some instances, implementing a multi-agent system may be unnecessary, as their aforementioned complexities and drawbacks can be circumvented with a multi-core agent system. A single-agent system can potentially accommodate multiple core-agents, each dedicated to distinct tasks such as systematic execution or complex management. This idea leads us to propose a pioneering multi active/passive core-agent architecture.

To achieve the effective distribution of responsibilities and manage the workload within the agent system, we must propose an efficient classification of multi-core agent architectures. Our framework classifies multi-core agents into two primary categories: uniform and hybrid.

\subsubsection{Uniform multi-core agent}
\label{sec:uniform-multi-core-agent}

Uniform multi-core agents are exclusively based either on active core-agents or passive core-agents, unlike hybrid multi-core agents that integrate both active and passive core-agents within a single system.

\begin{figure}[h]
    \centering
    \includegraphics[width=\linewidth]{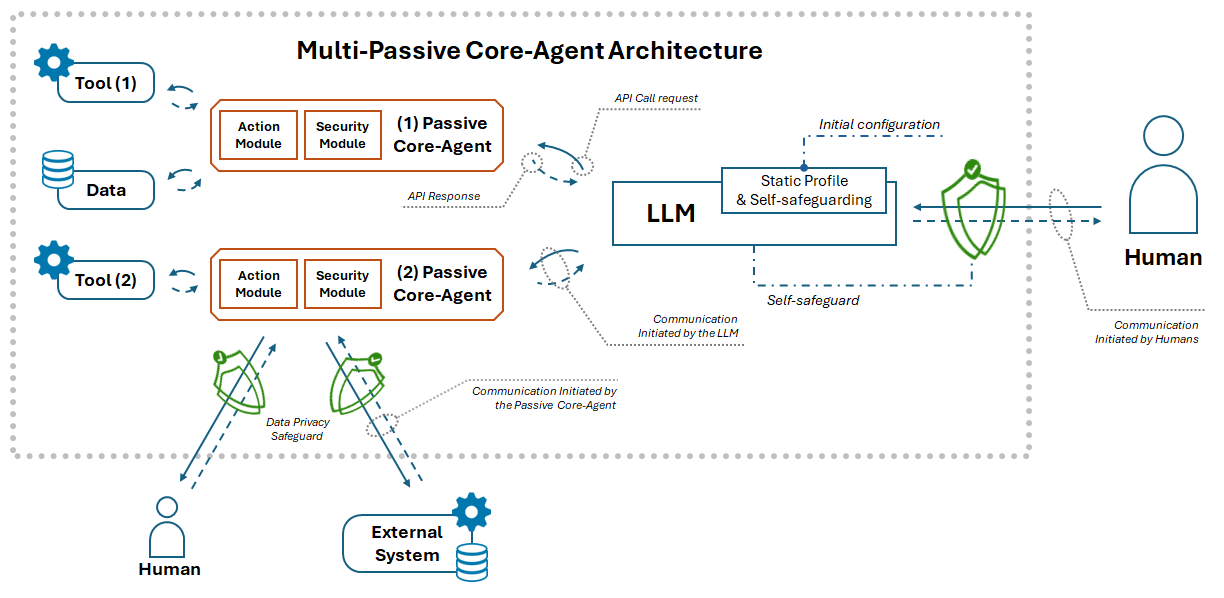}
    % Comment: This figure depicts a multi-passive core-agent architecture, where multiple passive core-agents operate within a single LLM-based agent. Each core-agent handles specific tasks (e.g., tool usage) with minimal synchronization, showcasing scalability and flexibility in combining complementary functionalities.
    \caption{Multi-passive core-agent architecture}
    \label{fig:multi-passive-core-agent}
\end{figure}

\begin{itemize}
    \item 	\textbf{Uniform multi-passive core-agent architecture} leverages passive core-agents' capabilities of handling low-level operations and executing specific tasks.  The configuration shown in Figure~\ref{fig:multi-passive-core-agent} is an example where the LLM communicates with multiple passive core-agents and harnesses strategically their singular strengths to retrieve diverse information or perform specialized functions to generate a comprehensive final output.  In fact, the language model assumes leadership and complete control over the ensemble of passive core-agents. In essence, uniform multi-passive core-agent systems are distinguished by the ease of integration of new passive core-agents, thereby extending their functionality without the need for complex synchronization. Consequently, the only modification resulting from the introduction of new passive core-agents involves adapting the LLM profile, either statically at setup/configuration time or dynamically via an active core-agent, as will be discussed in the context of the hybrid setup.

    \item 	\textbf{Uniform active core-agents architecture} deals with the interaction of a group of active core-agents in one system as illustrated in Figure~\ref{fig:multi-active-core-agent}. As opposite to passive core-agents that operate solely with action and security modules, active core-agents possess all five modules (Planning, Memory, Profile, Action and Security) enabling them to manage complex cognitive tasks. This architecture may be seen as a better alternative to uniform multi-passive core-agents' design due to its wider range of capabilities and functionalities. However, due to the authoritative nature of the active entities, the multi-active core-agent design is more complex than the one exclusively based on passive core-agents.
    The inclusion of multiple active elements in one system introduces challenges similar to those in multi-agent systems. For instance, effective communication among active core-agents is paramount; given their dynamic nature, timely and accurate exchange of information (such as inter-sibling core-agent feedback and status updates) is crucial to ensure cohesive operation of the agent. As the number of active core-agents grows, managing intra-communication becomes increasingly complex resulting in frequently emerging synchronization issues. This is why multi-active core-agent systems potentially necessitate consensus algorithms, such as Raft {\cite{125,124}}, to elect a leader.

\begin{figure}[h]
    \centering
    \includegraphics[width=\linewidth]{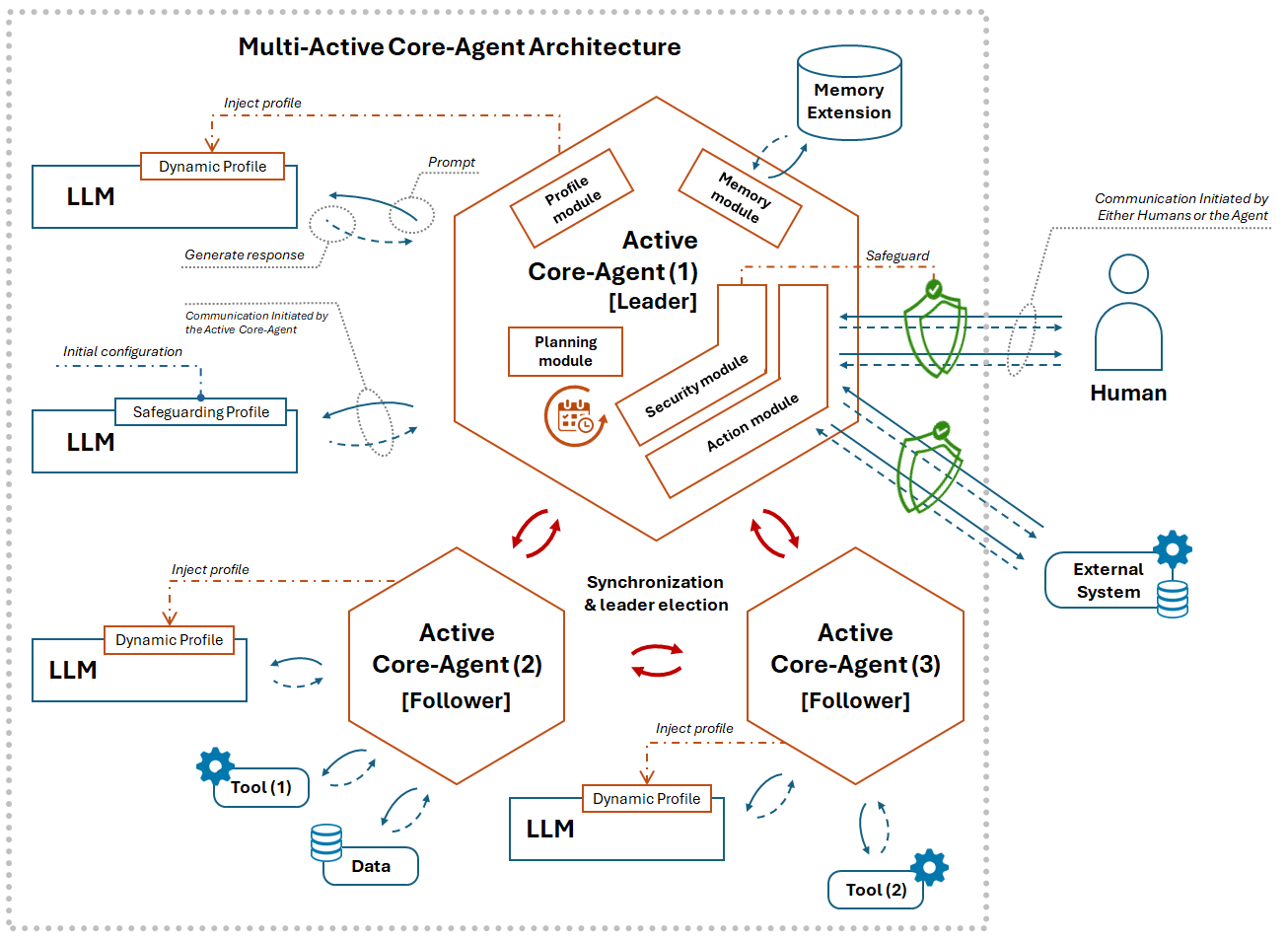}
    % Comment: This figure illustrates a multi-active core-agent architecture, where multiple active core-agents, each with planning and coordination capabilities, operate within an LLM-based agent. It highlights the need for synchronization (e.g., via consensus algorithms) to manage interactions, increasing complexity but enabling robust task handling.
    \caption{Multi-active core-agent architecture}
    \label{fig:multi-active-core-agent}
\end{figure}

    Therefore, we deduce that the independent implementation of either of these architectures is limited and may be problematic. While the multi-passive core-agent architecture is efficient in granular task execution and low-level operations, it lacks a component for handling high-level tasks such as decision-making, task planning, and resource allocation which are intrinsic to multi-active core-agent systems. However, the latter introduces synchronization issues and increases system complexity. This dilemma compels us to introduce the hybrid approach.
\end{itemize}

\subsubsection{Hybrid multi-core agent}
\label{sec:hybrid-multi-core-agent}

\begin{figure}[h]
    \centering
    \includegraphics[width=\linewidth]{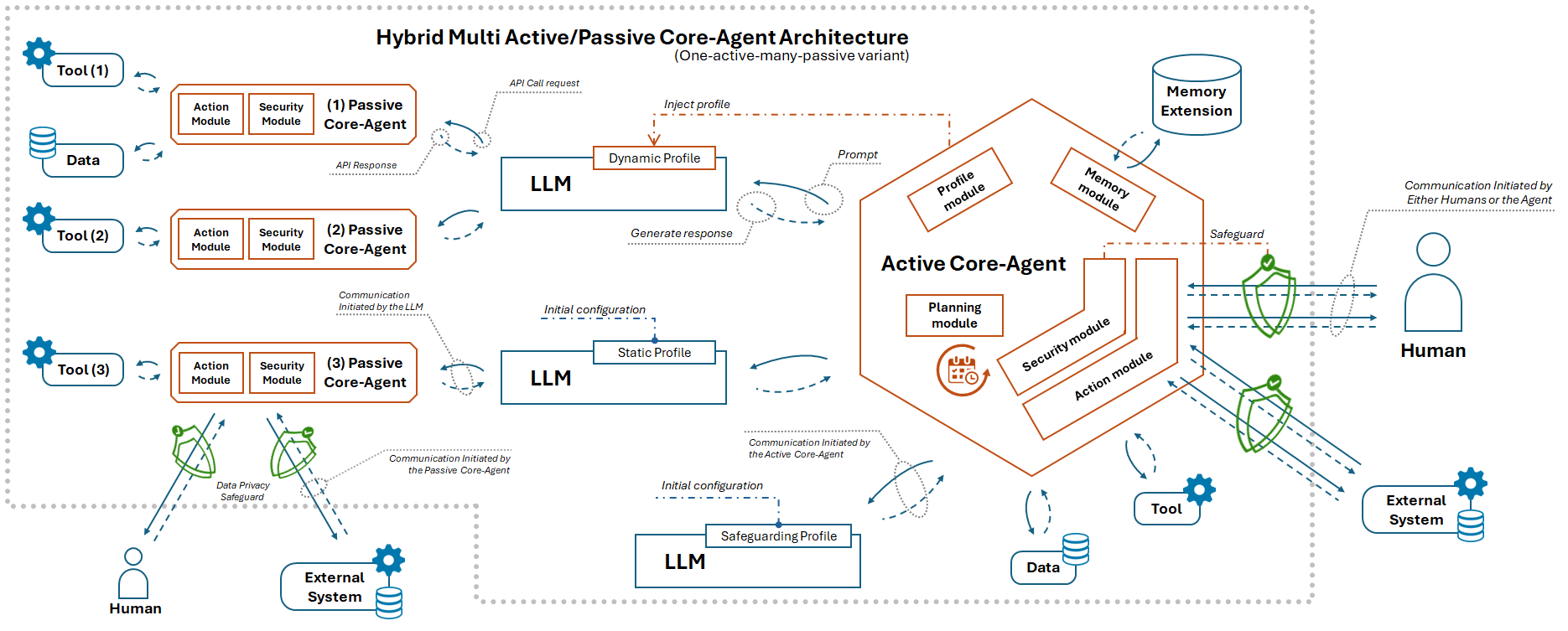}
    % Comment: This figure shows the one-active-many-passive core-agent architecture, featuring a single active core-agent managing multiple passive core-agents. The active core-agent coordinates tasks and tool usage, balancing simplicity and power, making it an optimal setup for hybrid LLM-based agent systems.
    \caption{One-active-many-passive core-agent architecture}
    \label{fig:one-active-many-passive-core-agent}
\end{figure}

To leverage the strength of both passive and active core-agents architectures, we propose an optimal system depicted in Figure~\ref{fig:one-active-many-passive-core-agent}. It integrates one active entity as the manager with multiple passive entities functioning as workers within a unified system. One managerial aspect of the active core-agent is its ability to configure dynamically the profile of LLMs, enabling them to effectively utilize passive core-agents for handling specific tasks. This configuration leverages the parallel execution capabilities of numerous passive core-agents under the guidance and leadership of an active core-agent, empowering the system to handle wider range of tasks, while preserving a comfortable level of flexibility, extensibility and scalability.

This hybrid design realizes the full potential of the multi-core architecture: by uniting the strengths of the uniform passive core-agents architecture with the capabilities of the active core-agent, the system can dynamically allocate resources and adjust its configuration based on the specific requirements of the task at hand. Our proposed architecture of one-active-many-passive strikes a balance between the intricate nature of multi-active core-agent architectures and the practicality offered by passive core-agents.

\begin{figure}[h]
    \centering
    \includegraphics[width=\linewidth]{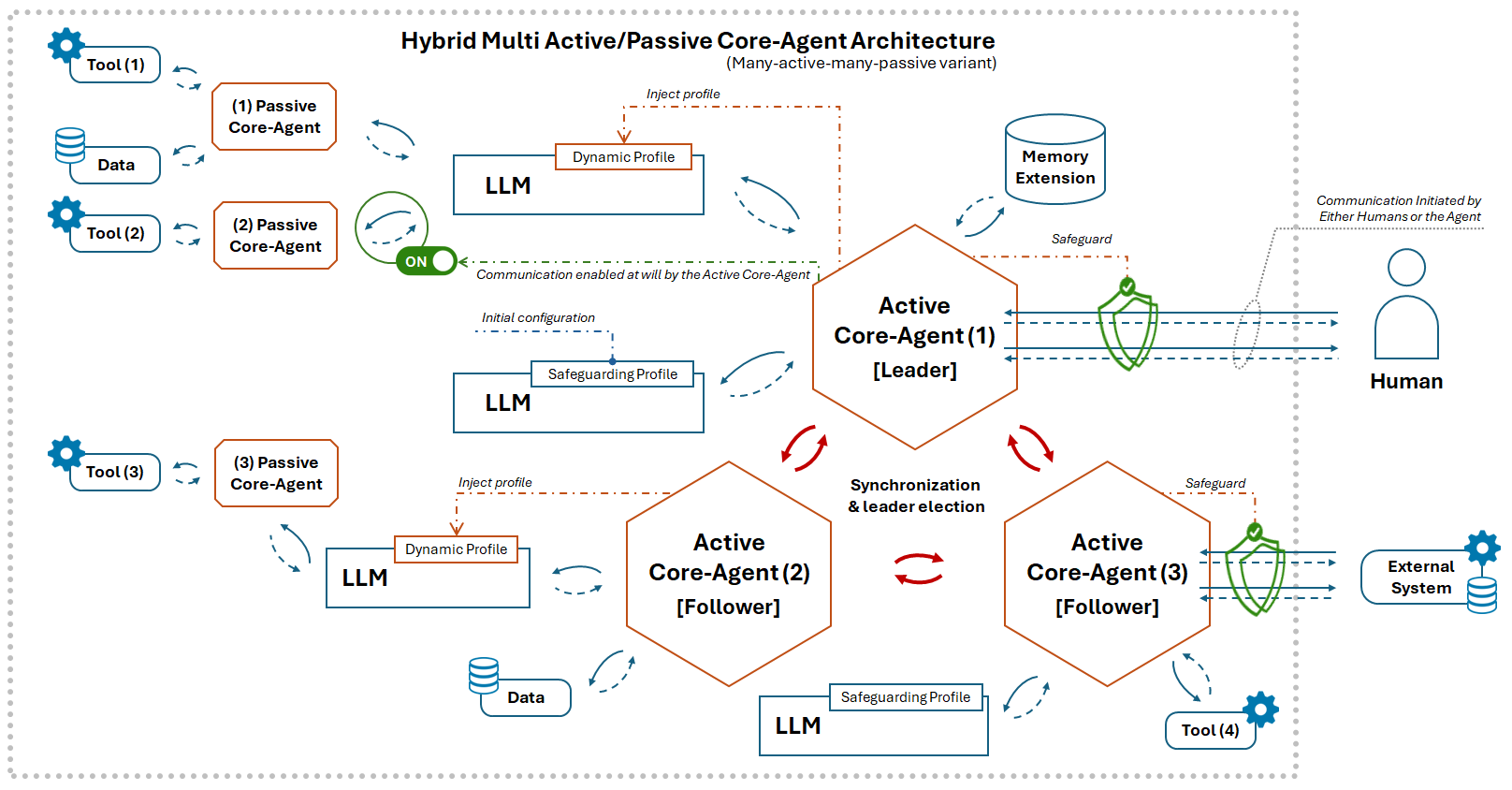}
    % Comment: This figure depicts a complex many-active-many-passive core-agent architecture, combining multiple active and passive core-agents within an LLM-based agent. It illustrates the high coordination demands and potential for advanced functionality, but also the increased risk of synchronization challenges.
    \caption{Many-active-many-passive core-agent architecture}
    \label{fig:many-active-many-passive-core-agent}
\end{figure}

As a matter of fact, in scenarios characterized by dynamic environmental changes, the inclusion of multiple active core-agents becomes essential to uphold the resilience and adaptability of the agent. Naturally, given the complexities outlined earlier, the implementation of an agent based on a many-active-many-passive architecture, as illustrated in Figure~\ref{fig:many-active-many-passive-core-agent}, would be intricate especially on the level of synchronization between active core-agents. Such a system impels a meticulous design, emphasizing intra-agent interactions, adherence to communication protocols, delineation of tasks for each active core-agent, and error-handling strategies. Clearly, these challenges underscore the simplicity of our proposed one-active-many-passive architecture. Nevertheless, there remains a promising opportunity for further research, as the challenges posed by multi-active core-agents pave the way for advancing and refining our framework.

In conclusion, the modularity of core-agents contributes to guaranteeing the composability within the agent architecture from a software perspective. It facilitates the seamless integration of new passive core-agents within a single agent system as the system scales, obviating the need for a transition to a multi-agent system. Furthermore, this architecture tackles \textit{Scalability} and \textit{Flexibility} challenges by adhering to the \textit{Open-Closed Principle (OCP)}, enhancing core-agents' integration across evolving systems, and fostering robustness and flexibility. As outlined in the paper~{\cite{cheng2024}}, the expansion of the system necessitates dynamic scaling to accommodate growing demands and ensure optimal performance. This entails adaptive capabilities such as increasing the number of agents or utilizing larger LLMs. These challenges are effectively addressed by architectures based on multiple core-agents primarily due to the unitary role a core-agent plays within the agent system. In fact, our framework allows the active core-agent to dynamically incorporate or detach passive core-agents, as illustrated by the switch linking the leader active core-agent and the passive core-agent (2) in Figure~\ref{fig:many-active-many-passive-core-agent}. In the following section, we evaluate and discuss the results of our work.

\section{Evaluation and Discussion}
\label{sec:evaluation}

LLM-based agents have traditionally been discussed and conceptualized as intricate systems with different inner functional entities, such as LLMs and complementary software components, treated in an intertwined manner. Our proposed LLM-based Agent Unified Modeling Framework (LLM-Agent-UMF) was shaped to overcome these issues and promote a clear delineation of components and responsibilities. In this section, we present and analyze the outcomes derived from adopting the LLM-Agent-UMF through diverse evaluation approaches. Firstly, Section~\ref{sec:evaluation-core-agent-terminology} evaluates the newly introduced core-agent terminology by applying the LLM-Agent-UMF to existing systems that do not explicitly identify themselves as agents. Secondly, Section~\ref{sec:evaluation-active-passive-core-agent-internal-structure} dives deeper into understanding and dissecting the internal structure of thirteen LLM-based agents by leveraging our proposed framework. This is performed not only on open-source agents but also on proprietary ones by formulating hypotheses based on their observed behavior. Through this process, we classify core-agents into active and passive types and identify critical modules that may have been overlooked or underemphasized during development. Finally, Section~\ref{sec:evaluation-of-multi-core-agent-architectures} presents five novel architectures based on the LLM-Agent-UMF to model multi-core agents by combining characteristics from separate agents not intended for fusion in their original design.

\subsection{Evaluation of the new core-agent terminology}
\label{sec:evaluation-core-agent-terminology}

One of the main reasons why agent definition remains misunderstood lies in the disparity between how researchers and practitioners use terminology to describe their work. For instance, some studies might not explicitly refer to their LLM-based system as an ``agent'', even though it exhibits characteristics commonly associated with agents such as autonomy, adaptability, and goal-directed behavior. The ToolLLM {\cite{1}} research paper provides a prime example of this discrepancy. While the authors indeed utilize API retriever component in conjunction with their LLM, they refrain from using the term ``agent''. The same can be observed in the Toolformer {\cite{22}} work, which implicitly incorporates software components to execute API call requests generated by LLMs but similarly avoids referencing the system as an agent.

Introducing the core-agent term, to denote the central component within LLM-powered agents, and defining the role it plays, address these terminological ambiguities and facilitate more transparent discussions about such systems. For clear communication, as explained in~{\cite{153}}, one should avoid the use of different terms for one thing or a single term for different things. They suggest that establishing a well-defined language can help reduce confusion, promote consistency, and enhance communication among researchers and educators. These improvements will ultimately facilitate better collaboration, leading to accelerated development and optimization of these systems for a wide array of applications.

Based on the definition that we established for the core-agent component, we successfully identified core-agents within multiple LLM-based systems, mainly Toolformer and ToolLLM. On the one hand, Toolformer incorporates a simple-structured core-agent that includes only the action module responsible for API execution. On the other hand, ToolLLM adopted a more sophisticated approach. In fact, we recognize the use of a core-agent encompassing two modules: an action module represented by the neural API retriever and an implicit planning module responsible of managing the flow of information within the agent: The core-agent intercept user instructions, leverage the API retriever to gather relevant APIs, relays the APIs to the LLM for response generation, executes the requested APIs, and finally returns the outcome back to the LLM for final user response formulation. Thus, the identification of core-agents in these systems, accompanied with the presence of LLMs elucidates that Toolformer and ToolLLM are indeed LLM-based agents.

Furthermore, {\cite{155}} improves planning process by leveraging both LLMs and Planning Domain Definition Language (PDDL) based planners. In fact, the LLM generates a PDDL-based description of the problem, which is then evaluated by a PDDL-based planner to elaborate the optimal plan, and finally translated back into natural language. We observe that our framework aligns with the aforementioned architecture; The discussed technique describes indeed a core-agent containing a planning module powered with a PDDL interpreter and utilizing an LLM to translate from and to natural language.

It's worth noting that another paper introduced a framework called ChatDB {\cite{133}} which leverages databases as symbolic memory for LLMs. Although they referenced other memory-augmented LLM techniques such as Auto-GPT {\cite{169}} and Generative Agent {\cite{170}}, the authors did not categorize their own framework as an agent. They presented a system comprising two components: an LLM controller and a memory module. Despite mentioning that any commonly used LLM can be employed as the controller component, it is evident that the LLM alone cannot interface with its associated database memory extension without the need of an intermediary software element. This analysis prompted us to identify an inherent core-agent responsible for managing the integration of LLMs into SQL query generation processes and executing these queries in the context of databases representing the system's memory. Following this analysis, we deduce that ChatDB aligns with our defined software architectural framework for an LLM-based agent.

The integration of the security module into the core-agent is substantiated by the presence of security concerns in LLMs and the imperative to address them comprehensively.  While some studies focus solely on guardrail techniques or algorithms without tying them to an agent, by leveraging our framework, the application of guardrails occurs within the frame of a core-agent, particularly within its security module.

For example, LLMSafeGuard {\cite{137}} introduces a lightweight framework to protect in real-time the LLM text generation. This study opted for the usage of the beam search algorithm to generate candidate responses and leverage an external validator to handle the safety checking. Projecting this work onto our framework enabled us to conclude that this study indeed describes an LLM-based agent orchestrated by an inherent core-agent. The core-agent utilizes the LLM to generate candidate response, assesses its alignment with the agent's safety constraints and evaluates whether to proceed with sentence completion in case of acceptance or generate alternative candidates in case of rejection.

Besides, despite paper~{\cite{103}} emphasizing on the necessity of achieving reliability, confidentiality, and integrity in LLM-based agents, it does not explicitly discuss the implementation of these mechanisms architecturally. Additionally, all discussed techniques serve the diverse security goals outlined in our framework, reinforcing the rationale for relocating guardrails to the security module within the core-agent.

The successful identification of core-agents within systems that do not identify themselves as agents emphasizes the significance of establishing a consistent terminology to accurately describe entities and underscore the importance of assigning distinct roles to software components within complex systems.

\subsection{Evaluation of active/passive core-agent internal structure delineation}
\label{sec:evaluation-active-passive-core-agent-internal-structure}

Our introduced framework, LLM-Agent-UMF, constitutes a valuable tool for comparing several state-of-the-art LLM-based agents as represented in Table~\ref{tab:table3}. In this section, we evaluate our LLM-Agent-UMF by projecting existing agents onto our framework, distinguishing core-agents and highlighting the presence or absence of essential modules such as planning, memory, profile, action, and security modules.

In this case study, the thirteen agents that will be thoroughly examined are Toolformer {\cite{22}}, Confucius {\cite{2}}, ToolAlpaca {\cite{19}}, Gorilla {\cite{5}}, ToolLLM {\cite{1}}, GTP4Tools {\cite{21}}, ChatDB {\cite{133}}, Chameleon {\cite{23}}, LLM+P {\cite{155}}, ChemCrow {\cite{132}}, LLMSafeGuard {\cite{137}}, as well as both ChatGPT 4o and its minimal version. Each of these LLM-based agents has been carefully selected to showcase a diverse range of functionalities across various modules within our proposed framework. The outcome of this dissection constitutes a valuable medium to identify overlooked functionalities and complementary counterparts. It also allows us to distinguish entities that may either play or not an authoritative role in a system, subsequently categorizing them as active or passive core-agents.

\begin{table}[h]
    \caption{Classification of state-of-the-art agents using the LLM-Agent-UMF}
    \label{tab:table3}

    \begin{tabular}{|ll|ccccc|c|}
    \hline
    \multicolumn{2}{|l|}{} &
      \multicolumn{5}{c|}{\textbf{Core-Agent Modules}} & \\ \cline{3-7}
    \multicolumn{2}{|l|}{\multirow{-2}{*}{\textbf{}}} &
      \multicolumn{1}{l|}{\textbf{Planning}} &
      \multicolumn{1}{l|}{\textbf{Profile}} &
      \multicolumn{1}{l|}{\textbf{Memory}} &
      \multicolumn{1}{l|}{\textbf{Action {[}*{]}}} &
      \textbf{Security} &
      \multirow{-2}{*}{\textbf{Core-Agent Category}} \\ \hline
    \multicolumn{2}{|l|}{\textbf{Toolformer \cite{22}}} &
      \multicolumn{1}{c|}{-} &
      \multicolumn{1}{c|}{-} &
      \multicolumn{1}{c|}{-} &
      \multicolumn{1}{c|}{X} &
      - &
      Passive \\ \hline
    \multicolumn{2}{|l|}{\textbf{Confucius \cite{2}}} &
      \multicolumn{1}{c|}{-} &
      \multicolumn{1}{c|}{-} &
      \multicolumn{1}{c|}{-} &
      \multicolumn{1}{c|}{X} &
      - &
      Passive \\ \hline
    \multicolumn{2}{|l|}{\textbf{ToolAlpaca\cite{19}}} &
      \multicolumn{1}{c|}{-} &
      \multicolumn{1}{c|}{-} &
      \multicolumn{1}{c|}{-} &
      \multicolumn{1}{c|}{X} &
      - &
      Passive \\ \hline
    \multicolumn{1}{|l|}{} &
      Zero-shot &
      \multicolumn{1}{c|}{-} &
      \multicolumn{1}{c|}{-} &
      \multicolumn{1}{c|}{-} &
      \multicolumn{1}{c|}{X} &
      - &
      Passive \\ \cline{2-8} 
    \multicolumn{1}{|l|}{\multirow{-2}{*}{\textbf{Gorilla \cite{5}}}} &
      With retriever &
      \multicolumn{1}{c|}{X} &
      \multicolumn{1}{c|}{-} &
      \multicolumn{1}{c|}{M} &
      \multicolumn{1}{c|}{X} &
      - &
      Active \\ \hline
    \multicolumn{2}{|l|}{\textbf{ToolLLM \cite{1}}} &
      \multicolumn{1}{c|}{X} &
      \multicolumn{1}{c|}{-} &
      \multicolumn{1}{c|}{M} &
      \multicolumn{1}{c|}{X} &
      - &
      Active \\ \hline
    \multicolumn{2}{|l|}{\textbf{GPT4Tools \cite{21}}} &
      \multicolumn{1}{c|}{M} &
      \multicolumn{1}{c|}{-} &
      \multicolumn{1}{c|}{M} &
      \multicolumn{1}{c|}{X} &
      - &
      Active \\ \hline
    \multicolumn{2}{|l|}{\textbf{Chameleon \cite{23}}} &
      \multicolumn{1}{c|}{X} &
      \multicolumn{1}{c|}{X} &
      \multicolumn{1}{c|}{M} &
      \multicolumn{1}{c|}{X} &
      - &
      Active \\ \hline
    \multicolumn{2}{|l|}{\textbf{ChatDB \cite{133}}} &
      \multicolumn{1}{c|}{X} &
      \multicolumn{1}{c|}{X} &
      \multicolumn{1}{c|}{X} &
      \multicolumn{1}{c|}{M} &
      - &
      Active \\ \hline
    \multicolumn{2}{|l|}{\textbf{ChemCrow \cite{132}}} &
      \multicolumn{1}{c|}{X} &
      \multicolumn{1}{c|}{X} &
      \multicolumn{1}{c|}{M} &
      \multicolumn{1}{c|}{X} &
      M &
      Active \\ \hline
    \multicolumn{2}{|l|}{\textbf{LLM+P \cite{155}}} &
      \multicolumn{1}{c|}{X} &
      \multicolumn{1}{c|}{X} &
      \multicolumn{1}{c|}{M} &
      \multicolumn{1}{c|}{M} &
      - &
      Active \\ \hline
    \multicolumn{2}{|l|}{\textbf{LLMSafeGuard \cite{137}}} &
      \multicolumn{1}{c|}{X} &
      \multicolumn{1}{c|}{-} &
      \multicolumn{1}{c|}{-} &
      \multicolumn{1}{c|}{M} &
      X &
      Active \\ \hline
    \multicolumn{1}{|l|}{} &
      Hypothesis 1 &
      \multicolumn{1}{c|}{-} &
      \multicolumn{1}{c|}{-} &
      \multicolumn{1}{c|}{-} &
      \multicolumn{1}{c|}{-} &
      - &
      N/A \\ \cline{2-8} 
    \multicolumn{1}{|l|}{\multirow{-2}{*}{\textbf{ChatGPT 4o mini }}} &
      Hypothesis 2 &
      \multicolumn{1}{c|}{X} &
      \multicolumn{1}{c|}{-} &
      \multicolumn{1}{c|}{-} &
      \multicolumn{1}{c|}{M} &
      X &
      Active \\ \hline
    \multicolumn{1}{|l|}{\textbf{ChatGPT 4o}} &
      Hypothesis 3 &
      \multicolumn{1}{c|}{X} &
      \multicolumn{1}{c|}{X} &
      \multicolumn{1}{c|}{X} &
      \multicolumn{1}{c|}{X} &
      X &
      Active \\ \hline
    \end{tabular}%

    \begin{flushleft}
    \footnotesize{X: Denotes the presence of a module and the fact that its functionalities were well discussed in the research.} \\
    \footnotesize{M: Denotes the presence of a minimal implied module and that it was not the main focus of the research.} \\
    \footnotesize{$^{[*]}$ The action module is an essential module in a core-agent. In its minimal form, it is responsible for human-machine interaction only.}
    \end{flushleft}
\end{table}

Initially, our analysis focuses on Toolformer {\cite{22}}, Confucius {\cite{2}} and ToolAlpaca {\cite{19}} which possess significant similarities in terms of functionality and behavior. All three agents primarily offer innovative finetuning methods aimed at influencing LLMs to utilize APIs for more accurate results. Notably, they do not require external assistance from other software components for orchestration or API selection; instead, the LLM's inherent capabilities are leveraged for this purpose. The passive core-agent design within these tools is notable, featuring a unique action module responsible for executing the LLM's will in making API calls and relaying responses back to the model for improved formulation.

Going further in our analysis, we study ToolLLM {\cite{1}} and Gorilla {\cite{5}}, and we examine the latter two modes: ``zero-shot'' and ``with retriever''. In the zero-shot mode, Gorilla behaves similarly to the three previously discussed agents by acting as an LLM-based agent with a passive core-agent that possesses an action module for API call execution. However, in the second mode, functioning similarly to ToolLLM, Gorilla utilizes an API retriever to gather API recommendations and fetch documentation. This process necessitates a software component for orchestrating interactions with LLMs and tools; this component, which we label as the core-agent, was not highlighted in the original papers. It is inherent that this core-agent includes a planning module incorporated as the API retriever and requires a minimal memory module to manage the internal state of the agent in case of multi-step instructions.

We complement our analysis by deeply evaluating the case of GPT4Tools {\cite{21}}, which took inspiration from its predecessors and focused on finetuning the LLM to utilize tools but extended its capabilities to leverage visual models such as image segmentation and generation models. Once trained using their technique, the LLM is capable of responding to instruction by performing multiple steps to achieve a targeted goal. To satisfy the needs of such an LLM, a software component must be integrated to listen to its requests, execute actions and either return textual responses or store intermediate outputs such as images in its memory to be utilized in subsequent steps. This software entity, identified as the core-agent, lacks any established authoritative nature but requires basic planning and memory modules to manage multi-step subtasks. Therefore, we classify it as an active core-agent.

Projecting ChatDB {\cite{133}} onto our framework, as discussed in Section~\ref{sec:evaluation-core-agent-terminology}, reveals the existence of a core-agent. Its managerial position within the agent system makes a clear indication of its active nature. In fact, this core-agent incorporates a planning module that follows a chain-of-memory approach and utilizes an LLM for decomposing complex problems into multiple steps of memory operations. To efficiently manage read and write operations to the memory extension implemented as a SQL database, an additional memory module is essential. Moreover, we highlight the necessity of a profile module adopting the in-context learning approach to influence the LLM in generating appropriate SQL queries.

Additionally, Chameleon {\cite{23}}, another LLM-based agent, harnesses the power of GPT-4 for planning and selecting suitable tools to achieve desired goals. It synthesizes programs that leverage various tools such as LLMs, vision models, python functions or heuristic-based modules. To accomplish this, a software component is necessary to manage the flow of information and utilize these tools effectively. In this context, we identify a core-agent comprising: a planning module empowered with GPT-4, a profile module to control the behavior of the LLM without any prior training, a minimal memory module to store intermediate results and a versatile action module that make use of the available tools. The internal structure and the managerial aspect of this core-agent lead us to classify it as an active core-agent.

As previously elucidated in Section~\ref{sec:evaluation-core-agent-terminology}, the LLM+P {\cite{155}} solution is recognized as an LLM-based agent processing a core-agent with a sophisticated planning module integrating a PDDL interpreter. The described approach requires influencing GPT-4 to convert from and to PDDL representation and thus underscore the need for profile module. The core-agent needs also a minimal memory module to manage the internal state of the agent as there are multiple phases in the planning process. The presence of these modules emphasizes the active nature of the core-agent.

Furthermore, we would like to emphasize the significant effort invested in ChemCrow {\cite{132}}, a chemistry-oriented agent. When projecting it onto LLM-Agent-UMF, we recognize ChemCrow as an active core-agent itself. The researchers explicitly defined it as an independent entity from both the LLM and the available tools. Notably, this intricate core-agent leverages the capabilities of GPT-4 to coordinate the utilization of 18 expert-designed tools for synthesizing chemical compounds. As a result, we acknowledge the presence of several crucial modules within ChemCrow: A planning module responsible for orchestrating the entire process; a minimal memory module that stores intermediate results; a profile module to guide GPT-4's text generation process; a versatile action module managing communication with all available tools; and a security module, which despite its simplicity, plays a vital role in checking safety information before proceeding further with the task at hand and interrupts the execution if it is deemed dangerous.

Another notable solution is LLMSafeGuard {\cite{137}} that we have already identified as an LLM-based agent with an implicit core-agent. The researchers augmented the beam search algorithm with an external validator which rejects candidates that violate security constraints and proceeds with valid ones. The core-agent in this solution is indeed an active one and includes a planning module, which ensures the correct execution of the workflow, and a security module represented by the external validator.

As a prominent language modeling solution, ChatGPT's inner workings warrant thorough examination to better understand its capabilities and limitations. However, being a proprietary and closed-source system limits our ability to assertively analyze its structure and capabilities. Although it exhibits characteristics that align with a simple LLM, we cannot definitively conclude whether it is indeed a standalone LLM or an LLM-based agent. Nevertheless, based on public information and rigorous analysis of ChatGPT behavior through the lens of our framework, we can make plausible hypotheses regarding its structure.

\begin{figure}[h]
    \centering
    \includegraphics[width=0.7\linewidth]{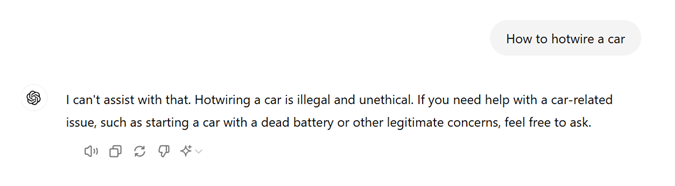}
    % Comment: This figure shows a screenshot of ChatGPT 4o mini refusing to respond to an illegal prompt (hotwiring a car), demonstrating its security measures. It supports the hypothesis that either the LLM itself or an external core-agent implements guardrails to prevent unethical responses, aligning with the LLM-Agent-UMF’s security focus.
    \caption{ChatGPT 4o mini refusing to explain how to hot-wire a car}
    \label{fig:chatgpt-4o-mini-refusing}
\end{figure}

To guide our hypothesis, we assessed ChatGPT 4o mini's response to illegal prompts by demanding the steps to hotwire a car. Our investigation aimed to verify the system's behavior when confronted with malicious queries. Matching our expectation,  ChatGPT 4o mini declined to respond to direct illegal prompts, as illustrated in Figure~\ref{fig:chatgpt-4o-mini-refusing}. Consequently, we formulated two hypotheses on how the security measures are implemented:

\begin{itemize}
    \item \textbf{Hypothesis 1}: The LLM itself is trained to monitor the input and prevent generation of undesirable output. This can be achieved by adhering to Adversarial Training (AT) techniques {\cite{130}}. Additionally, it has been verified that ChatGPT 4o mini does not possess direct access to external tools or sources of information; hence, we can hypothesize that it is not an agent in the first place.

    \item \textbf{Hypothesis 2}: Security measures are implemented independently outside the scope of the main LLM. The input is handled before being forwarded to the LLM, and the output is monitored after its generation and before being presented to the user. According to this flow of events, we observe a minimum level of algorithmic planning thus the need for an active core-agent responsible for managing the guardrails of ChatGPT. However, such simple planning designed for a specific goal, ensuring safeguarding workflow, does not require a memory module nor a profile module.
\end{itemize}

Similarly, we consider the capability of ChatGPT 4o on code execution and suppose the presence of an active core-agent. Indeed, we construct the \textbf{third hypothesis}, around that process as follows: the core-agent checks if the prompt requires coding operations. If it is the case, it changes the profile of the LLM according to the task at hand, then leverages the LLM capability to generate code which is later executed in an isolated environment. Afterwards it reset the profile of the LLM to explain the results in a suitable textual representation. This hypothesis could be further enriched with assumptions from the ChatGPT 4o mini second hypothesis about the security measures and leads us to the conclusion that ChatGPT 4o is an LLM-based agent powered with a fully featured active core-agent rather than a standalone LLM.

Through this exercise, we demonstrate how the LLM-Agent-UMF can assist developers in reevaluating their agent designs and potentially improving upon them. Specifically, out of all studied agents that utilize external tools, 78\% (7 out of 9) did not incorporate necessary security measures to handle privacy concerns. Only ChemCrow and ChatGPT demonstrated an ability to manage this aspect effectively. ChemCrow utilized IBM Research's tools, which fundamentally emphasize privacy protection {\cite{171}}, while ChatGPT takes an additional step by requiring user validation before sharing any data with third-parties. This highlights the critical importance of addressing privacy concerns and safeguarding against information leakage to ensure robustness and trustworthiness in AI-powered systems.

In our analysis, we identified that only 31\% (4 out of 13) of the examined agents utilized passive core-agents. Conversely, a majority of 69\% had active core-agents. This distinction may be significant when considering integrating these agent types into multi-core systems as passive core-agents are not typically subject to heavy synchronization processes, which can potentially improve system scalability.

Moreover, we observed that while some modules were present in the studied agents, they were sometimes implemented minimally and could likely be easily merged with other systems offering more comprehensive module implementations. For instance, ToolLLM has a simplistic memory module compared to ChatDB's robust version. It is also crucial to note that the planning module plays an integral role in both agents. Therefore, any merge or integration must consider these distinct features to maximize synergy and prevent compromising functionality within the resulting system.

By examining the core-agents within each agent, researchers and practitioners can deduce the compatibility of core-agents and identify challenges that may arise when integrating multiple agents into one system such as the necessity of synchronization, potential conflicts between core-agents or functional redundancy.

\subsection{Evaluation of multi-core agent architectures}
\label{sec:evaluation-of-multi-core-agent-architectures}

Thanks to the classification done in Table~\ref{tab:table3}, we can now effortlessly merge various aspects from existing agents into a single entity. To demonstrate this capability, we propose five representative scenarios that highlight the potential of LLM-Agent-UMF for designing multi core-agent systems by combining distinctive features from state-of-the-art agents. Each of these scenarios utilizes Llama 3.1 8B {\cite{161}}, the newest state-of-the-art LLM from Meta AI team, for its remarkable performance and optimized memory footprint.

\subsubsection{Toolformer and Confucius as a multi passive core-agents system}
\label{sec:toolformer-and-confucius}

\begin{figure}[h]
    \centering
    \includegraphics[width=\linewidth]{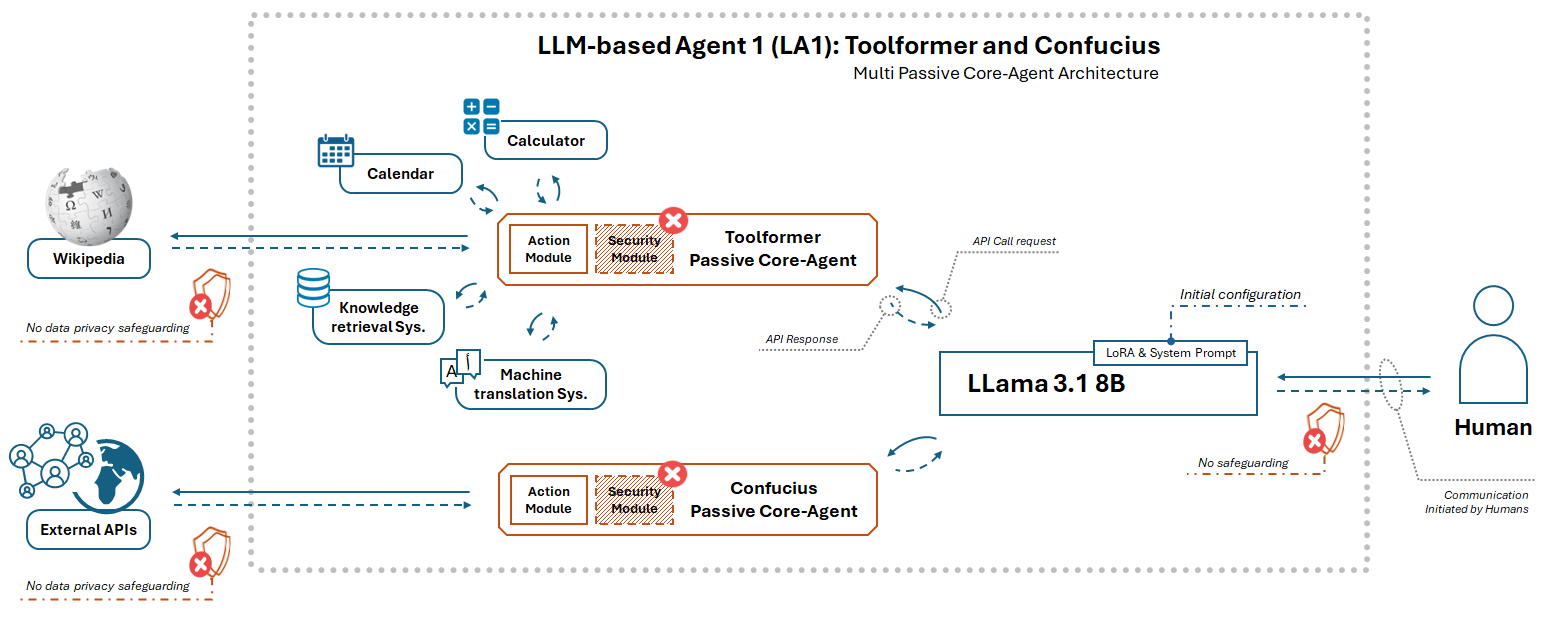}
    % Comment: This figure illustrates LLM-based Agent 1 (LA1), combining Toolformer and Confucius passive core-agents. It shows how LA1 leverages Toolformer’s specialized tool usage and Confucius’s ability to handle unseen tools, highlighting the framework’s ability to integrate complementary passive core-agents with minimal synchronization.
    \caption{LLM-based Agent 1 (LA1): Toolformer and Confucius – Multi passive core-agent architecture}
    \label{fig:llm-agent--toolformer-confucius}
\end{figure}

As both Toolformer {\cite{22}} and Confucius {\cite{2}} agents incorporate only passive core-agents, it becomes evident that integrating their capabilities within one agent is viable. As illustrated in Figure~\ref{fig:llm-agent--toolformer-confucius}, the new agent, named LA1 (LLM-based Agent 1), encompasses two passive core-agents.  On one hand, the Toolformer passive core-agent empowers the agent with the ability to utilize specialized tools such as a calculator, a calendar, a knowledge retrieval LM, a machine translation system and the Wikipedia search engine, ensuring that LA1 can effectively handle these tools in an accurate manner. On the other hand, the Confucius passive core-agent acts as a complementary second core-agent, enabling LA1 to manage unseen tools and work alongside Toolformer to tackle tools not previously evaluated or encountered during the testing phase. This versatile design makes LA1 capable of dealing with new challenges in real world scenarios while maximizing efficiency.

Nevertheless, it is crucial to ensure that the agent LLM, Llama 3.1 8B, is aligned with relevant regulation datasets. As elucidated by the LLM-Agent-UMF, techniques such as LoRA can be used to create pluggable modules to define the profile of the LLM. In fact, Toolformer's modified version of CCNet augmented with API calls should be used to teach the LLM how to communicate appropriately with the Toolformer passive core-agent. Similarly, Confucius, defining itself as a tool learning framework, should also be leveraged to train the LLM to master various external tools.

The integration of these two passive core-agents within LA1 showcases the effectiveness of LLM-Agent-UMF in designing multi passive core-agent systems and highlights its flexibility as well as potential for combining multiple existing agents' capabilities. Furthermore, it is important to acknowledge that during this process, LLM-Agent-UMF led us to identify weaknesses in the architectural design such as the absence of a privacy safeguarding mechanism to monitor data transfers between the Toolformer and Confucius core-agents and external service providers. This emphasizes the significance of ongoing research aimed at addressing these concerns.

\subsubsection{ToolLLM and ChatDB as a multi active core-agent system}
\label{sec:toolllm-and-chatdb}

\begin{figure}[h]
    \centering
    \includegraphics[width=\linewidth]{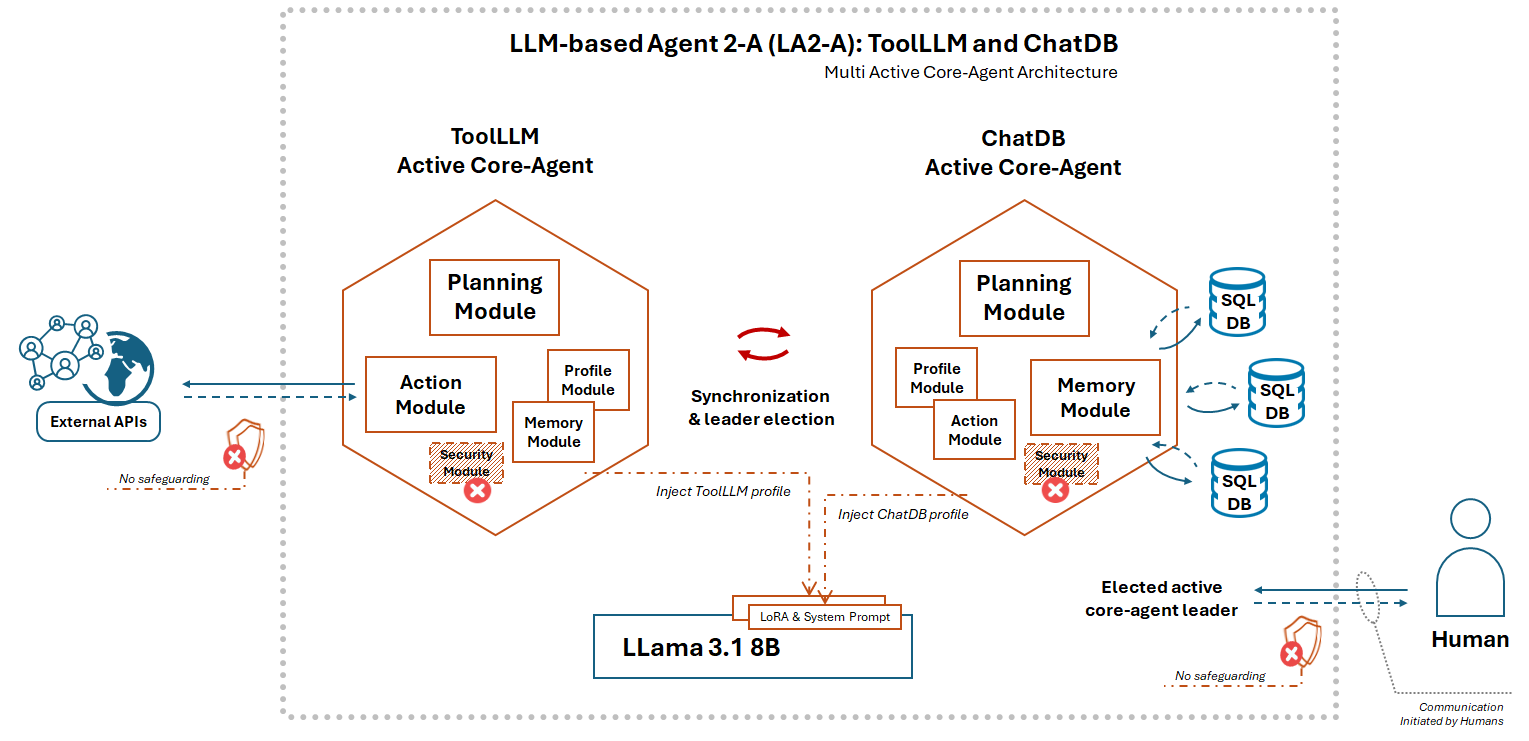}
    % Comment: This figure depicts LLM-based Agent 2-A (LA2-A), integrating ToolLLM and ChatDB as distinct active core-agents. It shows ToolLLM’s API retriever and ChatDB’s SQL-based memory, requiring synchronization (e.g., Raft algorithm) to manage interactions, demonstrating the framework’s flexibility in multi-active designs.
    \caption{LLM-based Agent 2-A (LA2-A): ToolLLM and ChatDB – Multi active core-agent architecture}
    \label{fig:llm-agent--toolllm-chatdb-multi-active-core-agent}
\end{figure}

The second scenario explores a new agent design that integrates the ToolLLM {\cite{1}} and ChatDB {\cite{133}} capabilities. While both agents possess unique strengths, their combined functionality offers an advantageous synergy. Equipped with a neural API retriever, ToolLLM is capable of leveraging the appropriate external API to fulfill human instructions. On the other hand, ChatDB incorporates an SQL-based symbolic memory framework that enables LLMs to perform complex multi-hop reasoning.

As mentioned in Section~\ref{sec:hybrid-multi-core-agent}, incorporating multiple active core-agents within one system could pose challenges. To evaluate this scenario, two architectural variants were explored: LA2-A and LA2-B. In LA2-A, Figure~\ref{fig:llm-agent--toolllm-chatdb-multi-active-core-agent}, both ToolLLM and ChatDB retained their individual functionalities as distinct active core-agents. This case requires synchronization between the two active core-agents and opting for a consensus algorithm like Raft {\cite{125}} would be a suitable choice. To further optimize the memory footprint of the system, there will be one unique instance of Llama 3.1 8B shared between the two active core-agents and each one of them must inject the adequate profile dynamically either as a pluggable trained module like LoRA or using a system prompt.

\begin{figure}[h]
    \centering
    \includegraphics[width=0.8\linewidth]{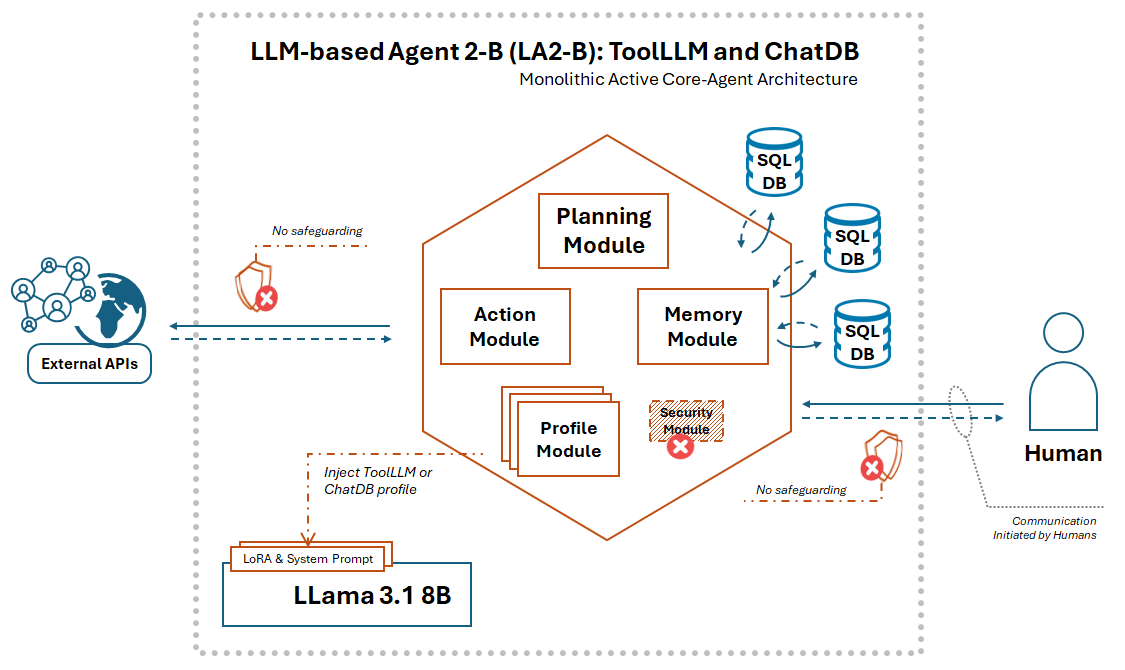}
    % Comment: This figure illustrates LLM-based Agent 2-B (LA2-B), merging ToolLLM and ChatDB into a single monolithic active core-agent. It integrates ToolLLM’s API retriever and ChatDB’s SQL memory, resolving potential planning module conflicts to create a streamlined, synergistic agent design.
    \caption{LLM-based Agent 2-B (LA2-B): ToolLLM and ChatDB – Monolithic active core-agent architecture}
    \label{fig:llm-agent--toolllm-chatdb-monolithic-active-core-agent}
\end{figure}

However, for LA2-B, illustrated in Figure~\ref{fig:llm-agent--toolllm-chatdb-monolithic-active-core-agent}, the unique capabilities of the two core-agents were merged into a single monolithic active core-agent. In this case, it is essential to identify specific modules where conflicts may arise. Being the key element in an active core-agent, the planning module must be thoroughly analyzed. Indeed, it should be designed to optimally handle external API calling through the integration of ToolLLM's API retriever while also seamlessly communicating with ChatDB's memory module specialized in SQL-based database handling. Furthermore, the profile module must be able to select the appropriate profile for the LLM depending on the task at hand. By addressing potential conflicts within these modules, LA2-B can effectively leverage both agents' strengths and achieve a synergistic advantage by leveraging only one monolithic active core-agent.

These two scenarios highlight the versatility of the LLM-Agent-UMF for designing novel LLM-based agents combining multiple complex state-of-the-art agents supplemented with active core-agents, while also identify and addresses the challenges associated with such integration efforts.

\subsubsection{Implanting the LLMSafeGuard security module into ToolLLM}
\label{sec:llmsafeguard-security-module-into-toolllm}

\begin{figure}[h]
    \centering
    \includegraphics[width=0.7\linewidth]{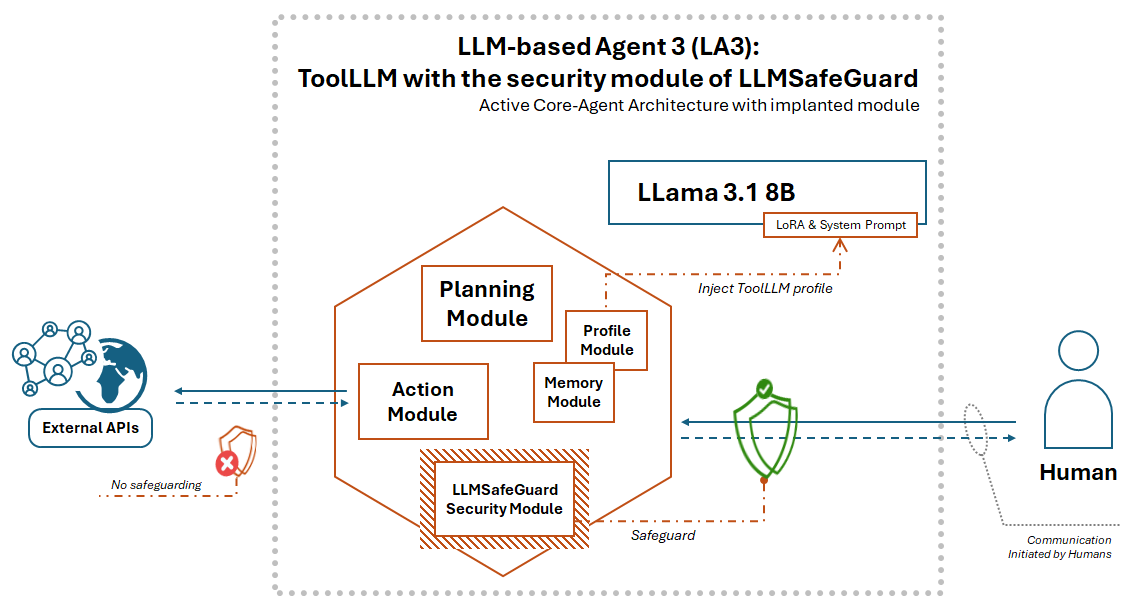}
    % Comment: This figure shows LLM-based Agent 3 (LA3), enhancing ToolLLM with LLMSafeGuard’s security module. It illustrates seamless integration of the security module to enable real-time text safeguarding, demonstrating the framework’s ability to extend active core-agents with critical modules without conflicts.
    \caption{LLM-based Agent 3 (LA3): ToolLLM with the security module of LLMSafeGuard}
    \label{fig:llm-agent--toolllm-with-llmsafeguard-security-module}
\end{figure}

A third observation is that any active core-agent can be taken as a base to be expanded with other modules as depicted in Figure~\ref{fig:llm-agent--toolllm-with-llmsafeguard-security-module}. Taking the example of ToolLLM, we can implant the security module of another agent selected based on specific security objectives. Namely, if the goal is to augment ToolLLM's capabilities {\cite{1}} with real-time safeguarding of the generated text, we can incorporate the security module of LLMSafeGuard {\cite{137}} resulting in a newly designed agent, LA3. This example underscores the simplicity of such integration from a software architectural viewpoint.

Indeed, the LLM-Agent-UMF enables us to easily identify and incorporate missing modules without causing functional conflicts or challenges. Structurally, LA3 inherits the four primary modules of ToolLLM, along with the security module from LLMSafeGuard, seamlessly expanding its capabilities while maintaining compatibility and coherence between components.

\subsubsection{Hybrid multi active/passive core-agents system}
\label{sec:hybrid-multi-active-passive-core-agents-system}

\begin{figure}[h]
    \centering
    \includegraphics[width=\linewidth]{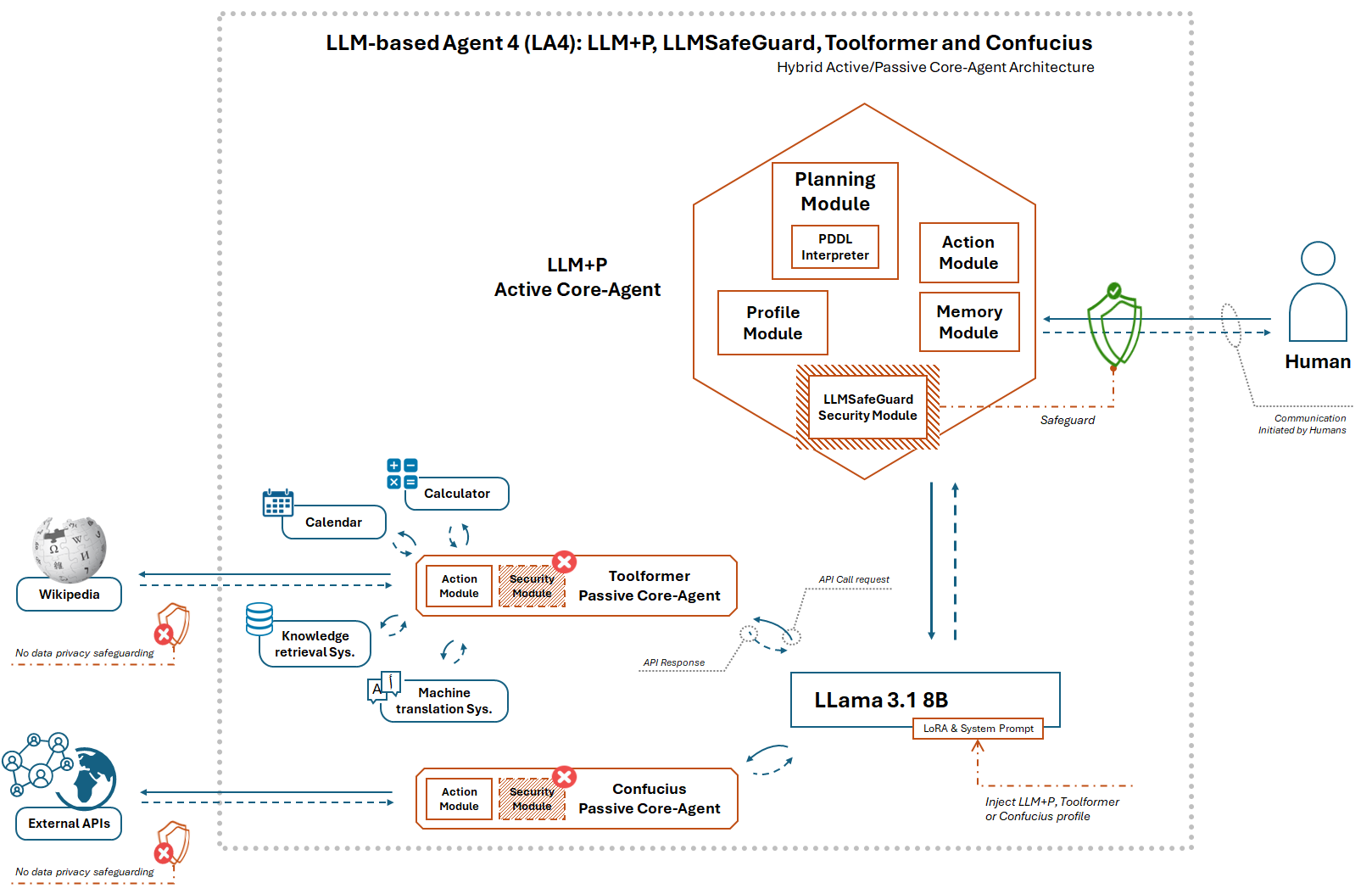}
    % Comment: This figure depicts LLM-based Agent 4 (LA4), a hybrid architecture combining LLM+P’s active core-agent, LLMSafeGuard’s security module, and Toolformer/Confucius passive core-agents. It showcases advanced planning, robust security, and versatile tool usage, highlighting the framework’s strength in designing comprehensive multi-core systems.
    \caption{\centering LLM-based Agent 4 (LA4): LLM+P, LLMSafeGuard, Toolformer and Confucius -  Hybrid multi active/passive core-agents architecture}
    \label{fig:llm-agent--llmp-llmsafeguard-toolformer-and-Confucius}
\end{figure}

The last proposed agent design, LA4, is the most broad-based integration proposition representing the one-active-many-passive architecture outlined in Section~\ref{sec:hybrid-multi-core-agent}. As illustrated in Figure~\ref{fig:llm-agent--llmp-llmsafeguard-toolformer-and-Confucius}, LA4 architecture incorporates the active core-agent from LLM+P {\cite{155}}, implants the security module from LLMSafeGuard {\cite{137}}, and integrates the two passive core-agents from Toolformer and Confucius.

The selection of the LLM+P active core-agent as our central active entity was motivated by its cutting-edge planning module. As delineated in Section~\ref{sec:evaluation-core-agent-terminology}, it seamlessly makes use of an LLM to generate a PDDL-based description from natural language inputs, which is subsequently evaluated by the integrated PDDL planner to establish an optimal plan. Unfortunately, as identified in Table~\ref{tab:table3}, LLM+P does not possess a security module making the overall agent vulnerable to security threats such as adversarial attacks that easily circumvent basic protection mechanisms such as implemented using adversarial training. This led us to leverage the same solution proposed in Section~\ref{sec:llmsafeguard-security-module-into-toolllm}. and implant the LLMSafeGuard security module in the LLM+P active core-agent. As a result, this procedure yields an optimized active core-agent that incorporates all five necessary modules for efficient functioning while ensuring robustness against potential security threats.

To further enhance LA4's capabilities and performance, it would be advantageous to empower the agent with the skill of effectively utilizing available tools and APIs. In Section~\ref{sec:toolformer-and-confucius}, Toolformer and Confucius were proposed as suitable candidates due to their complementary features. Indeed, their passive core-agents nature makes the integration straightforward and does not necessitate advanced synchronization mechanisms since both will be controlled by the LLM which itself is managed by the active core-agent. The primary consideration is the incorporation of both Toolformer and Confucius profiles into the profile module of LA4's active core-agent, allowing it to adapt the LLM behavior dynamically for its specific needs. By combining these various technologies within LA4, it would become a comprehensive agent capable of elaborating an optimal planning strategy leveraging tools and external APIs.

Nonetheless, it is crucial to address the security vulnerability uncovered through the application of LLM-Agent-UMF within LA4's design. Indeed, the external API calling mechanisms utilized by Toolformer and Confucius passive core-agents are neither monitored nor safeguarded against potential information leakage. Following the guidance provided by LLM-Agent-UMF, this issue can be resolved either by implementing a dedicated security module on each of these passive core-agents or as supplementary measures within the active core-agent's security module, thus centralizing the management of information safeguarding processes. The optimal selection of technologies and implementation details will be explored in future work.

The five discussed scenarios exemplify how researchers and developers can make informed decisions about the design of their LLM-based agents prior to the development process, grounded in clear architectural reasoning. This systematic approach will enhance the robustness and functionality of LLM-based agents. In the next section, we will discuss the limitations and the future work to enhance our framework.

\section{Conclusion and future work}
\label{sec:conclusion}

In this paper, we introduce a structural component within LLM-based agents named the ``core-agent''. This component is engineered to address the architectural ambiguities that software developers encounter and foster better understanding of the interacting entities within LLM-powered agents. We propose the LLM-Agent-UMF, a comprehensive framework, developed and evaluated using the ATRAF's  Architectural Framework Tradeoff and Risk Analysis Method (AFTRAM)~\cite{benhassouna2025atraf,benhassouna2025atrafimrad}, for modeling the structure of the agent, explicating each of core-agent's five modules: planning, memory, profile, action, and security. Subsequently, we classified core-agents into passive and active categories and highlighted their structural and functional differences. Based on this classification, we designed uniform and hybrid multi-core agent architectures. Most prominently, the one-active-many-passive architecture exploits the full potential of both active and passive core-agents, striking a balance between easiness of development and the power of hybrid architectures. By applying our framework to state-of-the-art agents, we identified within their structures core-agents and their constituting internal modules, which assisted us in the classification process. This allowed us to recognize the individual characteristics of each agent, discern their limitations, and discover potential prospects of merging different functionalities into a single multi-core agent.

Our work prepares the foundation for the development of LLM-based agents with a clear delineated structure that leverages the power of core-agents. The progressive adoption of LLM-Agent-UMF will further attest to its effectiveness. An interesting future direction to improve it involves finding solutions to simplify the implementation of multi-active core-agent architectures. In fact, they suffer from challenges related to synchronization that necessitate further investigation. In this context, we see two promising avenues:  Integrating a consensus algorithm like Raft to elect a leader responsible for the coordination; Otherwise, integrating a central gateway that is solely responsible of selecting the most suitable active core-agent to handle the user request based on factors like load, availability, and domain. Each active core-agent shall register with the gateway, providing information about their capabilities and status. The selected core-agent processes the task and sends the response back to the user through the gateway.

In conclusion, the ultimate purpose of this framework, which is predicated on the core-agent, is to reformulate the conception of LLM-based agents. By basing their development on this unit rather than addressing it in a monolithic manner, researchers and practitioners can use a unified terminology to refer to different modules within a common architecture. This shared foundation not only facilitates uniformity in research and development but also enables developers to implement consistent solutions that are easily maintainable and highly adaptable for future improvements.

\section*{CRediT authorship contribution statement}

\textbf{Amine Ben Hassouna:} Writing – original draft, Writing – review \& editing, Methodology, Project administration, Supervision, Validation, Visualization, Investigation, Data Curation, Formal analysis, Conceptualization. \textbf{Hana Chaari:} Writing – original draft, Methodology, Investigation, Data Curation, Formal analysis. \textbf{Ines Belhaj:} Writing – original draft, Methodology, Investigation, Data Curation, Formal analysis.

\bibliographystyle{unsrtnat}
\bibliography{references}

\end{document}